\documentclass[%
 amsmath,amssymb,
 aps,
 twocolumn,
 prb
]{revtex4-2}

\usepackage{graphicx}
\usepackage{dcolumn}
\usepackage{bm}
\usepackage{lipsum}
\usepackage{tikz}
\usetikzlibrary{arrows.meta}
\usetikzlibrary{decorations.pathmorphing}
\usepackage{braket}
\usepackage{comment}
\newcommand{\Vod}{V_{\mathrm{od}}}
\newcommand{\Vd}{V_{\mathrm{d}}}

\newcommand{\im}{\mathrm{i}}
\usepackage{hyperref}
\usepackage{dsfont}

\begin{document}

\title{Engineering Photon-mediated Long-Range Spin Interactions in Mott Insulators}

\author{Paul Fadler}
\affiliation{Department of Physics, Friedrich-Alexander-Universit\"at Erlangen-N\"urnberg, D-91058 Erlangen, Germany}
\author{Jiajun Li}
\affiliation{Paul Scherrer Institute, Condensed Matter Theory, PSI Villigen, Switzerland}
\author{Kai Phillip Schmidt}
\affiliation{Department of Physics, Friedrich-Alexander-Universit\"at Erlangen-N\"urnberg, D-91058 Erlangen, Germany}
\author{Martin Eckstein}
\affiliation{Department of Physics, University Hamburg, D-22607 Hamburg, Germany}

\date{\today}

\begin{abstract}
  We investigate the potential to induce long-range spin interactions in a Mott insulator via the quantum electromagnetic field of a cavity.
  The coupling between light and spins is inherently non-linear, and occurs via multi-photon processes like Raman scattering and two-photon absorption/emission with electronically excited intermediate states.
  Based on this, two pathways are elucidated:
  (i) In the absence of external driving, long-range interactions are mediated by the exchange of at least two virtual cavity photons.
  We show that these vacuum-mediated interactions can surpass local Heisenberg interactions in mesoscopic setups such as sufficiently small split-ring resonators.
  (ii) In a laser-driven cavity, interactions can be tailored through a hybrid scheme involving both external laser photons and cavity photons.
  This offers a versatile pathway for Floquet engineering of long-range interactions in macroscopic systems.
  In general, the derivation of these interactions requires careful consideration:
  Notably, we demonstrate that a simple phenomenological approach, based on a spin-photon Hamiltonian that captures Raman and two-photon processes with effective matrix elements, can be used only if the cavity is resonantly driven.
  Outside of these narrow resonant regimes as well as for the undriven case, a fourth-order series expansion within the underlying electronic model is necessary, which we perform to obtain long-range four-spin interactions in the half-filled Hubbard model.
\end{abstract}
\maketitle

\section{Introduction}\label{sec:intro}

Light is a unique tool to manipulate the properties of matter.
By subjecting matter to the time-periodic action of a strong laser pulse, one obtains Floquet Hamiltonians with topological band structures, artificial gauge fields, or modified super-exchange interactions~\cite{Bukov2015, Eckardt2017, Basov2017, delaTorre2021}.
An even more versatile approach to control materials would emerge with the potential to engineer long-range interactions.
Without back-action of matter on the driving field, Floquet engineering with classical time-periodic drives does not induce interactions between disconnected parts of a system~\footnote{If two parts of a system are not connected, their time-evolution operator factorizes even in the presence of a classical driving field.
Hence also the Floquet Hamiltonian, which is the generator of the stroboscopic time evolution, does not link the two parts.}.
However, the prospect of engineering long-range interactions with light arises when the quantum electromagnetic field acts as a force mediator, and confined geometries are used to enhance the light-matter coupling~\cite{Schlawin2022}.
Possible settings involve coplanar cavities akin to those employed in~\cite{Jarc2022}.
Alternatively, one can consider two-dimensional materials coupled to surface plasmon (SP) modes.
For a large in-plane wave-vector, the SP is exponentially localized at an interface~\cite{Economou69}, such that its momentum-dependent coupling to a two-dimensional material can be controlled by adjusting the distance between the material and the interface~\cite{Lenk2022, Ashida2020,Ashida2023}.

Proposals have been made to induce ferroelectricity~\cite{Lenk2022, Ashida2020} or superconductivity~\cite{Schlawin2019,Sentef2018} with photon- or plasmon-mediated interactions.
In these instances, the long-range interaction results from a linear coupling of the field to dipole-active transitions in matter, and the exchange of a single boson.
A potentially more versatile approach would rely on a nonlinear mechanism, where transitions in matter are induced by Raman scattering of two photons (or plasmons), or two-photon absorption and emission.
Unlike the linear mechanism, this is not limited to dipole-active transitions, such that, e.g., the electric field can lead to magnetic interactions, or van der Waals interactions~\cite{Petrosyan2008} between non-polar atoms.
Furthermore, interactions which arise through the nonlinear mechanism can be controlled by external driving, when one of the two modes is replaced by the classical field of a laser.
For example, an interaction between two distant sites in matter can be induced by means of Raman scattering between the laser and the cavity mode at one atom, the propagation of the photon to a different atom, and the reverse scattering process.

The idea to utilize an external laser drive in combination with nonlinear processes in matter for the design of long-range interactions has proven successful in a wide range of settings related to synthetic quantum matter, including  cavity-mediated interactions in cold atom experiments~\cite{Mottl2012,Landig2016,Klinder2015} or ion traps~\cite{Kapale2005}.
Consequently, the question arises whether related driving protocols can be realistically extended to solid-state systems.
Intriguing proposals along these lines encompass  the control of orbital pseudo-spin interactions~\cite{Chiocchetta2021} and superconducting pairing interactions~\cite{Eckhardt2023,Gao2020} within a cavity, using a nonlinear mixing of laser drive and cavity mode via the diamagnetic light-matter interaction or via near-resonant electronic intermediate states.

A common and intuitive way to understand such interactions theoretically is to represent the light-matter interaction as an effective two-boson scattering vertex $g_{\rm eff}(b_1^\dagger+b_1)(b_2^\dagger+b_2)\,O$ between the two modes and an operator $O$ in matter.
The vertex $g_{\rm eff}$ can be measured (e.g., in a Raman scattering experiment in free space) or computed, and the interaction is then obtained to second order in $g_{\rm eff}$ by eliminating the intermediate photon states.
However, this approach is in general valid only if the driving satisfies certain near-resonance conditions.
In this aspect, there is a crucial distinction between real materials and synthetic quantum matter.
In the latter case, one works with high quality cavities, and relies on a very well-defined separation of energy scales between the manifold of low-energy states targeted for the design of long-range interactions, the driving frequency, and the intermediate state.
For instance, cavity experiments with cold atoms employ lasers operating at optical frequencies to engineer long-range interactions in the MHz range~\cite{Landig2016}.
The enhancement of the interaction relies on a near resonance between the laser drive and the cavity, which both are sufficiently detuned from dipolar transitions of the atoms.
In contrast, in the context of condensed matter systems, the force mediator can be lossy (such as for a surface plasmon modes close to metallic interfaces), and the intermediate state can lie within a broad absorption band (e.g., it can be an electronic excitation in a dispersive band).
In this case, near-resonant driving protocols potentially lead to strong laser heating.
Similar to conventional Floquet engineering, the driving could therefore eventually suppress, rather than enhance, the collective orders which are supported by the induced interactions~\cite{Murakami2017}.

It is therefore necessary to understand light-induced interactions away from a driven resonance, or potentially in the undriven case where only virtual photons are exchanged.
In this paper, we showcase the relevance of off-resonant contributions to the interaction for long-range spin-interactions in the Fermi-Hubbard model.
In the Mott insulating limit at half filling, one can project out charge excitations from the Hubbard model to construct an effective Heisenberg spin Hamiltonian, while retaining the classical and/or quantum light-matter coupling~\cite{Mentink2015,Claassen2017,Sentef2020,Li2020,Li2022,Bostroem2022,Kiffner2019}.
The resulting nonlinear interaction can be viewed as a two-photon two-magnon scattering process with an effective vertex $g_{\rm eff}$.
The long-range interactions are then of fourth order in the fermionic hopping and lead to correlated spin flips at distant sites.
We explain how to derive this long-range interaction by degenerate perturbation theory of order four and thus evaluate how the more straightforward second-order perturbation theory in $g_{\rm eff}$ fails far away from the resonance between the laser drive and the cavity, and in the undriven case.
Moreover, we estimate that these interactions (even the vacuum-mediated ones), can be relevant for the collective behavior of light-matter hybrid systems under realistic conditions.

This work is structured as follows:
In Sections~\ref{sec:hub_in_cav},~\ref{sec:hub_basis} we explicitly introduce the
general setting and the cavity-coupled Hubbard Hamiltonian. 
Following that, we give an overview of the effective low energy description of this model in Section~\ref{sec:low_energy}.
We discuss the induced long-range interactions in Section~\ref{sec:lr_all}, where we first give general overview over their general structure in Section~\ref{sec:lr_overview} and then show how to derive them using a fourth order series expansion in Section~\ref{sec:fourthO}, or a second order scheme based on the effective spin-photon Hamiltonian in Section~\ref{sec:SpPh}.
We give an overview over the results for the undriven setting in Section~\ref{sec:results_closed}, paying special attention to where and why the two approaches deviate.
The driven setting is discussed in Section~\ref{sec:results_driven} with a focus on resonantly driving the cavity.
Finally, in Section~\ref{sec:real_systems} we relate to realistic experimental settings, in particular coupling a one band Mott insulator to a single-mode split-ring resonator.
An outlook is given in Section~\ref{sec:concl}.


\section{Model}\label{sec:model}

\subsection{Hubbard model in a cavity}\label{sec:hub_in_cav}

The specific setting which will be investigated in this paper is a Mott insulator, coupled to the quantum electromagnetic field, such as that of a cavity, and an additional classical time-periodic laser field.
It is described by the Hubbard Hamiltonian
\begin{align}
  H &= -t_0\sum_{\langle i,j \rangle,\sigma} c^{\dagger}_{i,\sigma}c^{\phantom{\dag}}_{j,\sigma} e^{\im \phi_{ij}}
    + U \sum_i n_{i\uparrow}n_{i\downarrow} + H_{\mathrm{field}},\label{eq:ham}
\end{align}
where $c^{\dagger}_{i,\sigma}$ ($c^{\phantom{\dag}}_{j,\sigma}$) create (annihilate) an electron with spin $\sigma\in\{\uparrow,\downarrow\}$ in a Wannier orbital at site $j$ (centered at position ${\mathbf{R}_j}$), $n_{i\sigma}=c_{i\sigma}^\dagger  c^{\phantom{\dag}}_{i\sigma}$, $U$ is the local Coulomb repulsion, and $t_0$ is the matrix element for hopping between nearest neighbours.
The coupling to the electromagnetic field is described within the dipolar representation, in which the vector potential $\mathbf{A}$ enters the Hamiltonian in the form of a quantum Peierls phase~\cite{LiGolez2020}
\begin{align}
  \phi_{ij}(t) &= q\int_{\mathbf{R}_i}^{\mathbf{R}_j}\mathbf{A}(\mathbf{r},t)\cdot d\mathbf{r},\label{peierls}
\end{align}
with the charge $q$ of the electron.
We will consider two distinct settings:
(i) The {\em isolated cavity}, in which only the quantum field $\mathbf{A}_{\rm qu}$ of the cavity is present, and (ii), the {\em driven cavity\/} in which the field contains a classical time-dependent laser field and the cavity field, $\mathbf{A}=\mathbf{A}_{\rm qu}+\mathbf{A}_{\rm cl}(t)$.
(We will use the terms quantum and cavity field interchangeably.)

For both settings we first make a number of simplifications:
For the cavity, we take into account only a single mode, such that the free field Hamiltonian is given by 
\begin{align}
H_{\rm field}=\omega_{\rm qu} a^\dagger a.
\end{align}
Moreover, we assume that both the laser field and the mode function are homogeneous over the sample with given polarization direction (unit vector $\hat n$), so that 
the vector potential can be represented as 
\begin{align}
 \mathbf{A}_{\rm cl}(t) &= \hat n A_0 \cos(\omega_{\rm cl}t),
\end{align}
for the classical laser field with amplitude $A_0$ and frequency $\omega_{\rm cl}$, and similarly
\begin{align}
 \mathbf{A}_{\rm qu} &= \hat{n} A_{\rm qu} (a^\dagger +a)\label{Aqu}
\end{align}
for the quantum field, where the field strength $A_{\rm qu}$ is controlled, e.g., by the mode volume of the cavity (see Section~\ref{sec:real_systems} for specific estimates).
Finally, we assume that the polarization direction $\hat{n}$ is parallel to the bonds $(i,j)$ of the lattice.
With this, the Peierls phase becomes 
 \begin{align}
  \phi_{ij} &= \xi_{ij} g_{\rm qu} (  a + a^\dagger) + \chi_{ij} g_{\rm cl}\cos(\omega_{\rm cl} t),\label{peierls2}
\end{align}
where $g_{\rm qu}=q|A_{\rm qu}\hat n \cdot (\mathbf{R}_i-\mathbf{R}_j)|$ and $g_{\rm cl}=q|  A_{\rm cl} \hat n\cdot (\mathbf{R}_i-\mathbf{R}_j)|$ are dimensionless coupling constants, and  $\xi_{ij}=\chi_{ij}=\pm1 $ for a bond in $\pm$ $\hat{n}$-direction.

Strictly speaking, the setting would apply for the specific case of a one-dimensional chain in a single-mode cavity, such as a split-ring resonator~\cite{Faist2014}.
The induced lang-range interactions in this case will be independent of distance (all-to-all).
Nevertheless, this setting contains all necessary ingredients to discuss the general role of off-resonant contributions to the interaction, which is the main purpose of this work.
Moreover, the expressions derived for the single mode case can relatively simply be extended to more general cases:
In particular, for the driven case an experimentally relevant setting would be if the ``cavity'' corresponds to a dispersive mode (such as a surface plasmon), such that the induced interactions acquire a nontrivial dependence on distance.

\subsection{Photon and Floquet basis}\label{sec:hub_basis}
For later reference let us state the expansion of the Hamiltonian in the photon and Floquet basis:
First, for the case without classical field, one can project the Hamiltonian onto a photon number basis $\ket{\nu}$.
The matrix elements  $H_{\mu\nu}=\bra{\mu} H \ket{\nu}$ become
\begin{align}
  H_{\mu\nu} &= \mathcal{H}^{0}_{\mu\nu} + \left(U\sum_i n_{i\downarrow}n_{i\uparrow} + \mu\omega_{\rm qu}\right)\delta_{\mu\nu},\nonumber\\
         \mathcal{H}^0_{\mu\nu} &= -t_0\sum_{\sigma}\sum_{\langle i,j\rangle} \im^{|\mu-\nu|}\xi_{ij}^{\mu-\nu}j_{\mu,\nu}(g_{\rm qu})c^{\dagger}_{i\sigma}c^{\phantom{\dag}}_{j\sigma},\label{eq:HamED}
\end{align}
with the function~\cite{Li2020}
\begin{equation*}
j_{\mu,\nu}(g_{\rm qu}) = e^{-g_{\rm qu}^2/2}\sum_{k=0}^\nu \frac{(-1)^k g_{\rm qu}^{2k+|\mu-\nu|}}{k!(k+|\mu-\nu|)!}\frac{\sqrt{\mu!\nu!}}{(\nu-k)!}
\end{equation*}
(for $\mu\geq \nu$ else the indices are swapped).

To discuss the driven system, where only the classical field is present, we can employ Floquet-theory. 
A Floquet state is represented in the extended Floquet Hilbert space, spanned by the matter Hilbert space and a discrete index $n\in\{0,\pm1,\pm2,...\}$, which will be called the sideband index or Floquet index in the following~\cite{Bukov2015,Eckardt2017}.
In the extended space, the Floquet states are determined with a time-independent Schr\"odinger equation, where the extended Hamiltonian takes the blockmatrix form $H^{mn}=  \delta_{mn} m\omega_{\rm cl} +  \tilde H_{m-n}$, with the Fourier components $\tilde H_{l}=\frac{1}{T}\int_0^T dt H(t) e^{\im l\omega_{\rm cl} t}$ of the $T$-periodic Hamiltonian ($T=2\pi/\omega_{\rm cl}$).
There is some freedom in choosing a Floquet gauge, which we use to obtain the same algebraic structure in the matrix elements as Eq.~\eqref{eq:HamED}.
For the present case, the Fourier transform of the classical Peierls phase gives
$\frac{1}{T}\int_{-T/2}^{T/2} dt e^{\im l \omega_{\rm cl} t } e^{\im \chi_{ij}g_{\rm cl} \cos(\omega_{\rm cl} t)}=\chi_{ij}^{l} \im^{|l|} J_{|l|}(g_{\rm cl})$,
with the Bessel function
$J_{n}(x)=\frac{1}{2\pi}\int_{-\pi}^\pi d\tau e^{\im n\tau + \im \sin(\tau)}=(-1)^n J_{-n}(x)$.
Hence,
\begin{align}
 H^{mn} &= \delta_{mn}\left( m\omega_{\rm cl} + U\sum_i n_{i\uparrow}n_{i\downarrow} \right)\nonumber\\
 &- t_0J_{|m-n|}(g_{\rm cl})\im^{|m-n|}
 \sum_{\left\langle i,j \right\rangle \sigma}\chi_{ij}^{m-n}c^{\dagger}_{i\sigma}c^{\phantom{\dag}}_{j\sigma}.\label{eq:mat_floq}
\end{align}

Finally, the driven case in the cavity can be considered as double expansion, in an extended Hilbert space spanned by the matter states, the Floquet index, and the photon number.
We will use greek indices to denote the occupation of the cavity and latin indices for the Floquet sidebands.
The matrix elements of the Hamiltonian in this basis are obtained as 
\begin{align}
 H^{mn}_{\mu\nu}
 = &\delta_{mn}\delta_{\mu\nu}\left( m\omega_{\rm cl} + \mu\omega_{\rm qu} + U\sum_i n_{i\uparrow}n_{i\downarrow} \right)\nonumber\\
 - &t_0\,\im^{|m-n|+|\mu-\nu|}J_{|m-n|}(g_{\rm cl})j_{\mu,\nu}(g_{\rm qu})\nonumber\\
  \cdot &\sum_{\left\langle i,j \right\rangle \sigma}\chi_{ij}^{m-n}\xi_{ij}^{\mu-\nu}c^{\dagger}_{i\sigma}c^{\phantom{\dag}}_{j\sigma}.\label{eq:mat_floq_cav}
\end{align}

The classical and the quantum field enter the kinetic part of the Hamiltonian in a similar way, but with one key difference:
Other than for the cavity case, the transition matrix elements in the Floquet case depend only on the difference $m-n$.
It is this block translational invariance which implies that the Floquet-Hamiltonian in a high-frequency expansion does not link disconnected parts of the lattice, which prevents drive mediated long-range interactions.

\subsection{Low-energy Hamiltonians}\label{sec:low_energy}

\subsubsection{Heisenberg model}\label{sec:low_energy_spin}
We consider the strong coupling limit $U \gg t_0$ of the Hubbard model at half filling.
Without coupling to the electromagnetic field, the leading order low-energy Hamiltonian is the Heisenberg Hamiltonian \mbox{$H_{\rm Hb}= J_{\rm ex}\sum_{\langle i,j\rangle} \vec{S}_i\cdot\vec{S}_j$} with antiferromagnetic exchange interaction $J_{\rm ex}=2t_0^2/U$, which is obtained by perturbatively eliminating charge fluctuations.
The purely laser-driven case ($g_{\rm qu}=0, g_{\rm cl}>0$) has been discussed extensively within Floquet theory:
In the Floquet block-matrix structure the effective Floquet spin model is obtained by eliminating perturbatively both the charge fluctuations and the Floquet sidebands.
The resulting Hamiltonian is a Heisenberg Hamiltonian with an exchange interaction $J_{\rm ex}^{F}(\omega_{\rm cl},g_{\rm cl})$~\cite{Mentink2015,Bukov2016,Itin2015}, which, depending on frequency and amplitude of the drive, can be positive (antiferromagnetic) and negative (ferromagnetic).
The reversal of the exchange interaction has been confirmed in cold gas experiments~\cite{Esslinger2017}.
Several generalizations have been discussed, including, e.g., higher order terms in $t_0/U$~\cite{Claassen2017}, higher order exchange processes via ligand orbitals~\cite{chaudhary2019}, orbital exchange processes~\cite{MuellerKK2023}, or doped states ($t$-$J$ model)~\cite{Gao2020b}.

In the opposite limit of a system only coupled to quantum photons, the Hamiltonian can be obtained likewise by perturbatively projecting out all charge and photon number fluctuations from some fixed cavity occupation $\nu$.
The resulting Hamiltonian can be considered as the exchange interaction $J_{\mathrm{qu},\nu}$ for a cavity with $\nu$ photons~\cite{Kiffner2019}.
One important limit is the empty cavity ($\nu=0$), which can be understood as a polaritonic dressing of the exchange interaction.
In the opposite limit, $\nu\to\infty$ with fixed $g_{\rm qu}\sqrt{\nu}=C$, one recovers the Floquet exchange Hamiltonian with amplitude $g_{\rm cl}=C$~\cite{Sentef2020}.

In real systems preparing stable multi-photon states $\ket{\nu}$ in the cavity are difficult to realize:
Dissipative processes lead to a finite lifetime of excitations and a linewidth broadening of $\Gamma$.
The effective Hamiltonian projected to a given occupation number can therefore in an undriven, open setting only describe the systems dynamics for $t\ll \Gamma^{-1}$.
Alternatively, these Hamiltonians can be understood as effective description of a driven dissipative system, where an external drive of the cavity mode stabilizes a Fock state $\ket{\nu}$ using the non-linearity of the hybrid cavity-matter system~\cite{Li2020}.
In this work we will, however, neglect dissipative processes and focus mainly on the induced dynamics by the empty cavity.

\subsubsection{Spin-photon Hamiltonian}\label{sec:sp_phot}
Alternative to eliminating both photonic and charge fluctuations, one can only eliminate the latter.
The resulting Hamiltonian is then defined on the subspace containing both spins and photons, and will be referred to as a spin-photon Hamiltonian.
To second order in $t_0/U$, it is given by~\cite{Sentef2020}
\begin{equation}
  H_{\rm SP} = J_{\mathrm{ex}}\sum_{\langle i,j \rangle}
  \mathcal{J}_{ij}( a^{\dagger},a ) \,P_{ij}^{S}
  + \omega_{\rm qu} a^{\dagger}a,\label{eq:spin-phot}
\end{equation}
where the interaction operator
\begin{equation}
\label{singletP}
P_{ij}^{S}=  \vec{S}_i\cdot\vec{S}_j - \frac{1}{4} 
\end{equation}
is the projector on a singlet on bond $(ij)$, and the exchange interaction is replaced by the operator $\mathcal{J}_{ij}(a^{\dagger},a)$.
Note that later on will only have one type of bond we will from now on drop the bond dependence in $\mathcal{J}$.
For the exact form of $\mathcal{J}$, see Ref.~\cite{Sentef2020} and App.~\ref{sec:OpForm}.
For a first understanding, and for later reference, we quote the leading order of the operator in $g_{\rm qu}$~\cite{Sentef2020}
\begin{align}
  \mathcal{J}\left(a^{\dagger},a\right) &= \mathcal{J}_0\left(a^{\dagger},a\right)
                         + \left( \mathcal{J}_2\left(a^{\dagger},a\right)a^2 + \mathrm{h.c.}\right),
  \end{align}
  with
  \begin{align}
  \mathcal{J}_0
  &= 1 - g_{\rm qu}^2 \frac{\bar{\omega}_{\rm qu}}{1+\bar{\omega}_{\rm qu}} + g_{\rm qu}^2 a^{\dag}a\frac{2\bar{\omega}_{\rm qu}^2}{1-\bar{\omega}_{\rm qu}^2} + \mathcal{O}(g_{\rm qu}^4),\label{eq:phot_diag_pert}
  \\
  \mathcal{J}_2
  &= g_{\rm qu}^2 \frac{\bar{\omega}_{\rm qu}^2 + 2\bar{\omega}_{\rm qu}^4}{(1-4\bar{\omega}_{\rm qu}^2)(1-\bar{\omega}_{\rm qu}^2)} + \mathcal{O}(g_{\rm qu}^4),\label{eq:phot_offdiag_pert}
\end{align}
and $\bar{\omega}_{\rm qu}=\omega_{\rm qu}/U$.
Taking matrix elements of $ \mathcal{J}( a^{\dagger},a )$ in a photon number state $\ket{\nu}$ would give the photon-dressed exchange interaction $J_{\mathrm{qu},\nu}=\bra{\nu}\mathcal{J}\ket{\nu}$ discussed in the previous section.
The off-diagonal terms in the photon number describe processes such as a spin flip (due to the operator $P_{ij}^{S}$) together with a change in the photon number, i.e.,
photon-magnon scattering.
For example, the two terms in the perturbative expression describe two photon absorption/emission ($ \mathcal{J}_2$) as well as photon scattering ($ \mathcal{J}_0$) on the spin system.

\begin{figure}
  \begin{center}
    \includegraphics{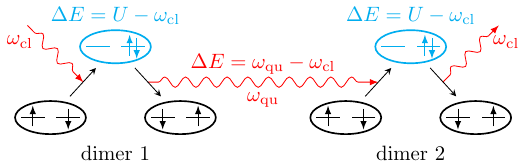}
  \end{center}
  \caption{
  Contribution to the resonant enhancement of the interaction at $\bar \Delta_{\rm qu}\ll1$, with two successive two-step processes:
  (i) A Raman process on dimer $1$: The system absorbs a photon from the classical drive during the first hopping on dimer $1$, leading to an intermediate virtual state with energy $U-\omega_{\rm cl}$, followed by a decay of the charge excitation under the emission of a cavity photon.
  (ii) The reverse Raman process on dimer $2$. The intermediate state energy between (i) and (ii) is ${\Delta_{\rm qu}=\omega_{\rm qu}-\omega_{\rm cl}}$.
   Note that if the laser drive is not resonant to $U$, also paths contribute to the leading order $1/\bar \Delta_{\rm qu}$, where the sequence of processes {\em within\/ }the same dimer is reversed (such as cavity photon emission before the laser photon absorption), but nevertheless the spin flips on the two dimers are successive.
    }\label{fig:resonant}
\end{figure}

Similarly, one can describe the driven cavity system by a mixed spin-photon-Floquet Hamiltonian, which is obtained from Eq.~\eqref{eq:mat_floq_cav} by perturbatively eliminating charge excitations while keeping both cavity and sidebands in a multi-block scheme, see App.~\eqref{eq:app_driv_spin_phot}.
This gives a spin Hamiltonian in the extended Floquet/photon space, with the blockmatrix structure,
\begin{align}
  (H_{\rm SP})^{mn}_{\mu\nu} &= \delta_{mn}\delta_{\mu\nu}(m\omega_{\rm cl} + \mu\omega_{\rm qu})\nonumber\\
                    &+ J_{\mathrm{ex}}\sum_{\left\langle i,j \right\rangle}
                    P_{ij}^S\,\,
                    \mathcal{J}^{mn}_{\mu\nu}(g_{\rm cl},g_{\rm qu}).\label{eq:mbot_spin_phot}
\end{align}
For explicit expressions of the matrix elements $\mathcal{J}^{mn}_{\mu\nu}$,  see Eq.~\eqref{eq:J_sp_driven} in the appendix.
The matrix elements $\mathcal{J}^{mn}_{\mu\nu}$ describe the absorbtion/emission of $m-n$ photons from the classical drive under the absorbtion/emission of $\mu-\nu$ cavity photons from/into the cavity at occupation $\nu$ through second-order processes.
For example, $\mathcal{J}^{01}_{10}$ describes scattering of a photon from the drive to the cavity (left half of Fig.~\ref{fig:resonant}), while $\mathcal{J}^{10}_{01}$ describes the opposite scattering process (right half of Fig.~\ref{fig:resonant}).
These terms are therefore the matrix elements for Raman scattering on the spin system~\cite{Shastry1990}.
For details on the relation to Raman scattering, see App.~\ref{sec:raman}.

\section{Induced long-range interactions}\label{sec:lr_all}

\subsection{Overview}\label{sec:lr_overview}

We now turn to the central topic of this paper, the derivation of cavity-induced long-range spin interactions. 
In general, to obtain a photon-dressed spin model, we work in a regime where driving and cavity frequencies are of the same order of $U$, such that $U$ can be considered as a common high-energy scale, and $t_0/U$ is treated as a small parameter.
Similar to the Heisenberg model (Sec.~\ref{sec:low_energy_spin}), the effective spin Hamiltonian for the driven or undriven cavity is obtained by projecting the system to a  subspace with no charge excitations (spin only), a given cavity occupation $\nu$ (such as $\nu=0$), and the $0$th Floquet sector, while keeping virtual excitations to the other sectors perturbatively in $t_0/U$.

In order to obtain long-range interactions, one will need to go to fourth order in $t_0/U$. To this order, virtual tunnelling processes lead to three different contributions in the effective Hamiltonian:
(i) Corrections to the nearest neighbour exchange interaction beyond $t_0^2/U$, (ii), short-range three-spin and four-spin interactions, which are restricted to connected clusters of the lattice, and (iii), long-range cavity-mediated interactions between bonds $(ij)$ and $(kl)$ which are not connected by a hopping process.
Since a spin triplet state on a bond $(ij)$ does not allow electron tunnelling and therefore does not couple to light, the long-range interaction between the bonds can involve only singlet states, and it can therefore be written in the form  $P^S_{ij}P^S_{kl}$, with the singlet projectors~\eqref{singletP}.
Hence, the effective Hamiltonian takes the general form
\begin{equation}
H_{\mathrm{eff}} = \sum_{\langle i,j \rangle}J_{\mathrm{Hb},ij}P_{ij}^S + \sum_{\langle i,j \rangle \langle k,l \rangle} K_{ij,kl} P^S_{ij}P^S_{kl} + \ldots,\label{eq:int_general}
\end{equation}
where the ellipsis $\ldots$ refers to the short-range three-spin and four-spin terms.
They can give rise to interesting physics (see, e.g., Ref.~\cite{Claassen2017} and~\cite{Bostroem2022}), but in this work we only focus only on the long-range interactions, which can have a significant qualitative effect on the behavior of the spin model.
To derive the interaction $K_{i_1j_1,i_2j_2}$ between disconnected bonds $(i_1,j_1)$ and  $(i_2,j_2)$ on a lattice to fourth order in $t_0/U$, it is sufficient to consider two isolated dimers.
This is because the interaction term $P^S_{i_1j_1}P^S_{i_2j_2}$ contains a spin flip on each dimer, which already requires two hoppings within each dimer.
Hence, to fourth order in $t_0/U$, no further virtual excited states can be generated on sites other than $(i_1,j_1)$ and $(i_2,j_2)$.
We can therefore restrict the following analysis on a $4$-site system which only contains the two isolated dimers.
The effective Hamiltonian then takes the form
\begin{equation}
H_{\mathrm{eff}} = 2J_{\mathrm{Hb}} ( P^S_{i_1j_1}+ P^S_{i_2j_2}) + 8K P^S_{i_1j_1} P^S_{i_2j_2},\label{eq:proj}
\end{equation}
where the additional prefactors come from the sum over sites in Eq.~\ref{eq:int_general}.
The overall scale of the long-range interaction $K$ will be $K_0\equiv 2t_0^4/U^3$, with a dimensionless prefactor depending on the cavity occupation $\nu$, the light-matter coupling $g_{\rm qu}$, the laser driving strength  $g_{\rm cl}$, and the  ratios $\bar \omega_{\rm qu}=\omega_{\rm qu}/U$
and $\bar \omega_{\rm cl}=\omega_{\rm cl}/U$,
\begin{equation}
\label{Kparametrization}
K/K_0 \equiv \kappa^{(\nu)}(g_{\rm qu},g_{\rm cl},\bar \omega_{\rm qu},\bar \omega_{\rm qu}).
\end{equation}

Within the four site model, the interactions can be simply read off the spectrum:
The Hamiltonian has eigenenergies $E=8K+4J\equiv E_{SS}$ when both bonds are in a singlet state, $E=2J\equiv E_{S}$ when only one bonds is in a singlet state, and $E=0$ when both bonds are in a triplet.
Hence the interaction $K$ is given by
\begin{equation}
8K=E_{\rm SS}-2E_{\rm S}.\label{Kspindimer}
\end{equation}
One can therefore numerically determine the interaction by solving the Hubbard model~\eqref{eq:ham} for the two dimers as follows:
Also on the Hubbard model, the eigenstates states can be classified as singlet (which now includes the doubly occupied configurations) or triplet on each bond.
We define the energy $E_{\rm SS}^{\nu,m}(g_{\rm qu},g_{\rm cl},t_0)$ as the energy of the state which is adiabatically connected to the state with zero charge excitations, two singlets, the $m$th Floquet sector, and $\nu$ photons at $g_{\rm qu}=g_{\rm cl}=t_0=0$.
Similarly, the energy  $E_{\rm S}^{\nu,m}(g_{\rm qu},g_{\rm cl},t_0)$ is defined for one singlet.
By comparing with~\eqref{Kspindimer}, the effective interaction is therefore obtained from the energy difference
\begin{equation}
\Delta E^{\nu,0} = (E^{\nu,0}_{\mathrm{SS}}-\nu\omega_{\rm qu})  - 2(E^{\nu,0}_{\mathrm{S}}-\nu\omega_{\rm qu}).\label{eq:edExtract}
\end{equation}
Finally, the function~\eqref{Kparametrization} can be extracted by numerically taking the limit
\begin{equation}
\kappa^{(\nu)}(g_{\rm qu},g_{\rm cl},\bar \omega_{\rm qu},\bar \omega_{\rm qu})
\stackrel{t_0\to0}{=}
\frac{\Delta E^{\nu,0}(g_{\rm qu},g_{\rm cl},\bar \omega_{\rm qu},\bar \omega_{\rm qu},t_0)}{8\cdot2t_0^4 /U^3 }.
\label{limitt04}
\end{equation}
We have used this approach to benchmark the analytical perturbative expressions obtained below.
In practice, we numerically diagonalize the Hamiltonian~\eqref{eq:HamED} (for the undriven cavity) or~\eqref{eq:mat_floq_cav} (for the driven cavity), with a sufficiently high cutoff in the photon number and Floquet index to converge the result (for details see App.~\ref{sec:basis}).
To obtain the limit~\eqref{limitt04}, we evaluate the exact spectrum for different $t_0$ and extract the series coefficient from a polynomial fit.

In the following subsections we derive the expressions for the interaction to leading order in $t_0/U$ by means of two different series expansions.
At first, we discuss a full fourth-order perturbation theory in $t_0/U$, which gives the exact result.
Secondly, we describe the approach based on the spin-photon Hamiltonian introduced in Sec.~\ref{sec:sp_phot}, which describes photon-matter scattering with matrix elements $\propto t_0^2$ and eliminates the photons.

\subsection{Fourth-order perturbation theory}\label{sec:fourthO}
\subsubsection{Schrieffer Wolff transformation}\label{SecSWT}
We aim to derive the effective Hamiltonian in the subspace which contains no charge excitations (doubly occupied sites and holes), and a given photon number and Floquet index $m=0$.
All other states are energetically off-resonant and will be eliminated.
Virtual transitions to the off-resonant states determine the effective Hamiltonian in the target energy space.
The general procedure is to find a unitary transformation that decouples the target space and the off-resonant states up to a given perturbative order.
In the rotated basis, we can then project out the off-resonant states, while the resulting Hamiltonian is the effective Hamiltonian in the target space.
There are many different techniques for obtaining this transformation~\cite{Takahashi1977, Schrieffer1966, Shavitt1980, Lowdin1962, MacDonald1990, Uhrig2000}.
Below, we follow the Schrieffer-Wolff transformation as defined by Loss et al.~\cite{Bravyi2011}.

The Schrieffer-Wolff transformation is formulated for a general Hamiltonian of the type
\begin{align}
H=H_0+V,
\end{align}
where the unperturbed part $H_0$ does not mix target and off-resonant states, while the perturbation $V$ mixes the states.
We will later choose $H_0$ to contain the onsite interaction and cavity and sideband energies, and $V$ to be the hopping term.
The selection of the target space is accomplished through the choice of a projector $P_0$ (and its complement $P_1=1-P_0$), that projects onto the target space.
We further assume that there is an energy gap between the target space and the rest of the unperturbed Hilbertspace.
If we are able to diagonalize the unperturbed part of the Hamiltonian $H_0$, ($H_0\ket{j}=E_j\ket{j}$), we can obtain the effective Hamiltonian in the target space from a series expansion in $V$.
It is helpful to understand each application of a matrix element $\bra{i}V\ket{j}$ of the perturbation as step through the unperturbed Hilbertspace.
That way, we obtain the effective Hamiltonian as sum over all paths connecting the target space with itself.
The Schrieffer-Wolff transformation then determines the weights with which these path have to be summed up.
The weight of each path then depends on energy resolvents, which are conveniently expressed in terms of a resolvent superoperator.
For any operator $X$ with off-diagonal contribution $X_{\rm od}=P_0 X P_1 + P_1 X P_0$ the resolvent is defined as~\cite{Bravyi2011}
\begin{equation}
  \mathcal{L}(X_{\mathrm{od}}) = \sum_{i,j}\frac{\ket{i}\bra{i}X_{\mathrm{od}}\ket{j}\bra{j}}{E_i - E_j}.\label{eq:super_op}
\end{equation}
Here $\{\ket{i}\}$ and $\{\ket{j}\}$ thereby form complete eigenbases of different blocks of the unperturbed Hamiltonian $P_0H_0P_0$ and $P_1H_0P_1$.
To fourth order, the general expression reads
\begin{align}
  H_{\mathrm{eff}}
  &= P_0 H P_0 + \frac{1}{2}P_0[S_1,\Vod]P_0\nonumber\\
  &+ \frac{1}{2}P_0[\Vod,\mathcal{L}\left( [\Vd, S_1] \right)]P_0\nonumber\\
  &-\frac{1}{2}P_0[\Vod,\mathcal{L}\left( [\Vd, \mathcal{L}\left( [\Vd, S_1] \right)] \right)]P_0\nonumber\\
  &- \frac{1}{6}P_0[\Vod,\mathcal{L}\left( [S_1,[S_1,\Vod]] \right)]P_0\nonumber\\
  &-\frac{1}{24}P_0[S_1,[S_1,[S_1,\Vod]]]P_0,
  \label{longheff}
\end{align}
where 
\begin{align}
\label{eqs1generat}
S_1 = \mathcal{L}(\Vod)
\end{align}
is the leading order of the generator of the transformation~\cite{Bravyi2011}.

For our setting we can furthermore simplify the general expression:
At half filling, a hopping always creates a charge excitation.
Hence $P_0 \Vd = \Vd P_0 = 0$, and $\Vd$ vanishes in the low-energy subspace.
Additionally, since we limit the geometry to two disconnected dimers, there are no processes with three hoppings connecting the charge excitation free target space with itself ($P_0 \Vod \Vd \Vod P_0 = 0$), so that the third order is vanishing.
In the Fermi-Hubbard model at half filling, this holds true for all odd orders of the perturbation theory for arbitrary lattices as long as they do not contain odd-sized loops.
Furthermore, the unperturbed Hamiltonian $H_0$ acts trivially on states in our target space, i.e. $P_0 H_0 P_0 = E_0 \mathds{1}$.
This property makes it possible to shift the superoperator $\mathcal{L}$ as $P_0\mathcal{L}(X)YP_0 = - P_0X \mathcal{L}(Y)P_0$~\cite{Bravyi2011}, which furthermore simplifies the general expression.
Using these simplifications, up to fourth order we find
\begin{align}
  H_{\rm eff}
  &= P_0 H_0 P_0 + P_0 S_1 \Vod P_0 + \frac{1}{8}P_0[S_1,[S_1,[S_1,\Vod]]]P_0\nonumber\\
  &+ P_0\mathcal{L}\left( S_1\Vd \right) \Vd S_1P_0.\label{eq:eff_ham}
\end{align}
This expression will now be evaluated explicitly for the isolated and undriven cavity, respectively.
In the main text we will mainly summarize the resulting analytical expressions and their structure, while derivations are shifted to the appendices.

\begin{figure*}
\includegraphics{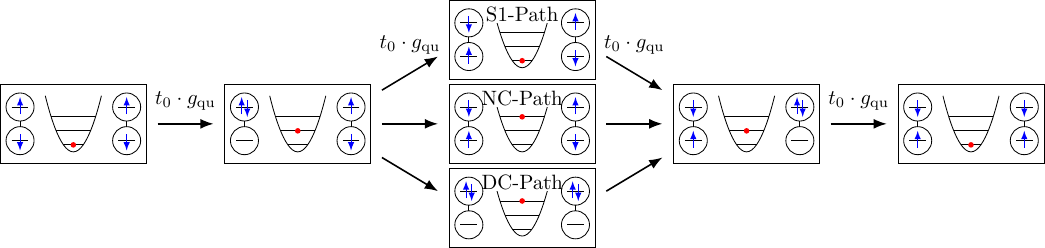}
\caption{
  Different representative paths through the unperturbed Hilbert space in order $g_{\rm qu}^4t_0^4$.
  The intermediate state can be from the low-energy part of the unperturbed Hilbertspace (S1-path) or the high-energy part.
  In the latter case, we can either have a cavity excitation with no charge excitations (NC-path) or two charge excitations and arbitrary cavity occupation (DC-path).}\label{fig:pert_sketch}
\end{figure*}

\subsubsection{Isolated cavity}
Let us fist discuss the isolated and empty cavity, i.e., we choose the low-energy part of the unperturbed Hilbert space with $\nu=0$ photons as target space.
This amounts to the projector
\begin{equation}
P_0=\prod_i(1-n_{i\uparrow}n_{i\downarrow})\otimes\ket{0}\bra{0}.\label{eq:proj_empty}
\end{equation}
Any hopping from the target space is part of $\Vod$, since it has to create a charge excitation.
For the diagonal part $\Vd$, which only acts outside of the target space, there are two possibilities:
(i), transitions between states with zero and one charge excitations but with the cavity in an excited state, or (ii) transitions between states with one and two charge excitations.

Using these processes/steps it is possible to create three different types of operator products/paths
which connect the target space with itself (see Fig.~\ref{fig:pert_sketch}):
The first ones we label as S1-paths, where only the off-diagonal part of the perturbation is responsible for the steps.
They are therefore described by the threefold nested commutator of Eq.~\eqref{eq:eff_ham}.
For those also depending on the diagonal part of the perturbation $\Vd$, we distinguish between those with \textbf{N}o \textbf{C}harge excitations in the intermediate state as the NC-paths and those with a \textbf{D}ouble \textbf{C}harge excitation in the intermediate state as the DC-paths.

After some algebra (see App.~\ref{sec:FourthODer}), we obtain the fourth-order terms of the Hamiltonian as a sum over the three contributions $  H^{(0)}_{\rm S1}+  H^{(0)}_{\rm NC}+  H^{(0)}_{\rm DC}$,
\begin{align}
 H^{(0)}_{\rm S1}
& =
 4K_0  \big(  P^S_{i_1j_1}\!+\!P^S_{i_2j_2}\!+\!2P^S_{i_1j_1}P^S_{i_2j_2}  \big)
  \!\!\!\!\sum_{\alpha,\beta,\gamma=0}^{\infty}  \!\!\!\!W^{0\alpha\beta\gamma0}_{\rm S1}
  \!\!\!\!,
\label{eq:s1}\\
 H^{(0)}_{\rm NC}
& =
 4K_0   \big(  P^S_{i_1j_1}\!\!+\!P^S_{i_2j_2}\!+\!2P^S_{i_1j_1}P^S_{i_2j_2}  \big)
\!\!\!\!\!\!\sum_{\alpha,\beta,\gamma=0}^{\infty} \!\!\!\!\!W^{0\alpha\beta\gamma0}_{\rm NC},
\label{eq:NC}
\\
 H^{(0)}_{\rm DC}
& =
 8K_0 P^S_{i_1j_1}P^S_{i_2j_2}
 \!\!\!\!\sum_{\alpha,\beta,\gamma=0}^{\infty} \!\!\!\!W^{0\alpha\beta\gamma0}_{\rm DC},
 \label{eq:DC}
\end{align}
where $W^{0\alpha\beta\gamma0}_{p}$ sums the contribution from all path of type $p\in\{$S1,NC,DC$\}$ with photon states $0\to\gamma\to\beta\to\alpha\to0$.
Explicit expressions for the path contributions are given by
\begin{align}
W^{0\alpha\beta\gamma0}_{\rm S1}&=\mathcal{J}^{0\alpha\beta\gamma0}(g_{\rm qu})\frac{\delta_{\beta0}(-1)^{\alpha+\gamma}  \left( 2 + (\gamma+\alpha)\bar{\omega}_{\rm qu} \right)}{\left( 1 + \alpha \bar{\omega}_{\rm qu} \right)^2 \left( 1 + \gamma\bar{\omega}_{\rm qu} \right)^2}
  \label{eq:s1path}
  \\
W^{0\alpha\beta\gamma0}_{\rm NC}&= -\mathcal{J}^{0\alpha\beta\gamma0}(g_{\rm qu})\frac{(1-\delta_{\beta0})(-1)^{\alpha+\beta+\gamma}(1+(-1)^{\beta})}{(1 + \alpha\bar{\omega}_{\rm qu})(\beta\bar{\omega}_{\rm qu})(1 + \gamma\bar{\omega}_{\rm qu})}
    \label{eq:NCpath}\\
W^{0\alpha\beta\gamma0}_{\rm DC}&=  -\mathcal{J}^{0\alpha\beta\gamma0}(g_{\rm qu})\frac{(-1)^{\beta}\left( 2 + (-1)^{\alpha+\gamma}(1 + (-1)^{\beta}) \right)}{(1 + \alpha\bar{\omega}_{\rm qu})(2 + \beta\bar{\omega}_{\rm qu})(1 + \gamma\bar{\omega}_{\rm qu})}
\label{eq:DCpath}
\end{align}
where
\begin{align}
\mathcal{J}^{\alpha\beta\gamma\delta\epsilon}(g_{\rm qu}) =& \im^{|\alpha-\beta| + |\beta-\gamma| + |\gamma-\delta| + |\delta-\epsilon|}
\nonumber\\
&\times \,\,\,j_{\alpha\beta}(g_{\rm qu})j_{\beta\gamma}(g_{\rm qu})j_{\gamma\delta}(g_{\rm qu})j_{\delta\epsilon}(g_{\rm qu}),\nonumber
\end{align}
and $K_0=2t_0^4/U^3$.
It is furthermore possible to extend this scheme to arbitrary cavity number states $\ket{\nu}$ by shifting the target space and using the projector
\begin{equation}
P^{(\nu)}_0 = \prod_i \left( 1 - n_{i\uparrow}n_{i\downarrow} \right)\otimes \ket{\nu}\bra{\nu}.
\end{equation}
For this case, one still has the same types of paths, but different amplitudes, phases and resolvents.
We find that this is accounted for by shifting the indices $\alpha,\beta,\gamma \rightarrow \alpha-\nu,\beta-\nu,\gamma-\nu$ everywhere but in the coupling amplitudes $j_{\mu\nu}$;
see App.~\ref{sec:fourthO_driven_deriv} and Eqs.~\eqref{eq:W_S1}-\eqref{eq:W_DC} for the contributions $H^{ (\nu)}_{\rm S1}$, $H^{ (\nu)}_{\rm NC}$, and $H^{ (\nu)}_{\rm DC}$, generalizing Eqs.~\eqref{eq:s1}-\eqref{eq:DC} to $\nu\neq 0$.

One can see that the fourth-order contribution contains both higher-order corrections to the exchange couplings (the terms proportional to $P^S_{i_1j_1}+P^S_{i_2j_2}$), and a mediated interaction (the terms proportional to the product $P^S_{i_1j_1}P^S_{i_2j_2}$).
By comparing Eqs.~\eqref{eq:s1}-\eqref{eq:DC} (or Eqs.~\eqref{eq:W_S1}-\eqref{eq:W_DC} for $\nu\neq0$) to Eq.~\ref{eq:proj} one can therefore immediately read off the long-range interaction in the parametrization of Eq.~\eqref{Kparametrization}.
We write
\begin{equation}
\label{kappa_sum}
\kappa^{(\nu)}(\bar{\omega}_{\rm qu},g_{\rm qu})
\equiv
\sum_{\mathrm{path}\in\{{\rm S1,NC,DC}\}}
\kappa^{(\nu)}_{\mathrm{path}}(\bar{\omega}_{\rm qu},g_{\rm qu}),
\end{equation}
where $\kappa^{(\nu)}_{\mathrm{path}}(\bar{\omega}_{\rm qu},g_{\rm qu}) = \sum_{\alpha,\beta,\gamma} W_{\mathrm{path}}^{\nu\alpha\beta\gamma\nu} $ is  the prefactor of the product $P^S_{i_1j_1}P^S_{i_2j_2} $ 
in the Hamiltonian  $H^{ (\nu)}_{\mathrm{path}}$.
In evaluating the sums, we introduce an upper cutoff $\nu_{\rm max}$ for the intermediate photon numbers $\alpha$, $\beta$, $\gamma$.
The result quickly converges with increasing $\nu_{\rm max}$~\cite{Li2020}.
The three contributions  $\kappa^{(\nu)}_{\rm S1}$, $\kappa^{(\nu)}_{\rm NC}$, and $\kappa^{(\nu)}_{\rm DC}$ will also be analyzed separately below.
For $\nu\neq 0$, these interactions come with an additional caveat to the one discussed in Sec.~\ref{sec:low_energy_spin}:
Close to resonances ($\mu\bar{\omega}_{\rm qu}=1$ for any $\mu\leq \nu$) the gap between the target space and the rest of the unperturbed Hilbertspace vanishes.
To ensure convergence of the perturbative series, this may limit the range of $t_0/U$ we can investigate~\cite{Bravyi2011}.
Luckily the contribution of high-order resonances requires many photon number transitions, and the convergence of the series is therefore controlled not only by $t_0/U$, but also by $g_{\rm qu}$.

\subsubsection{Driven Cavity}\label{sec:drivenFourthO}
Since the matrix elements of the driven cavity (see Eq.~\eqref{eq:mat_floq_cav}) have the same structure as those of the isolated cavity, the perturbation theory is very similar (see App.~\ref{sec:fourthO_driven_deriv}).
To account for the different Floquet-sidebands, we have to extend the projectors.
The target space is now defined by the doublon-free sector, a given cavity occupation $\nu$, and the zeroth Floquet sector:
\begin{equation}
P^{\nu}_0 = \prod_i \left( 1 - n_{i\uparrow}n_{i\downarrow} \right) \otimes \ket{\nu_{\rm qu}}\bra{\nu_{\rm qu}} \otimes \ket{0_{\rm cl}}\bra{0_{\rm cl}}.
\end{equation}
(Because of translational invariance in Floquet space, the effective Hamiltonian obtained by projection to Floquet sector $m$ is independent of $m$, and we choose $m=0$ without loss of generality.)
Analogous to Eq.~\eqref{kappa_sum}, the interaction can again be written as a sum of the contributions of S1-, NC- and DC-paths,
\begin{equation}
\label{kappa_sum_drivem}
\kappa^{(\nu)}(\bar{\omega}_{\rm qu},g_{\rm qu},\bar{\omega}_{\rm cl},g_{\rm cl})
\equiv
\!\!\!\!\!\!\!\!\!\!\sum_{\mathrm{path}\in\{{\rm S1,NC,DC}\}}\!\!\!\!\!\!\!\!\!\!
\kappa^{(\nu)}_{\mathrm{path}}(\bar{\omega}_{\rm qu},g_{\rm qu},\bar{\omega}_{\rm cl},g_{\rm cl}),
\end{equation}
where
\begin{equation}
\kappa^{(\nu)}_{\mathrm{path}}(\bar{\omega}_{\rm qu},g_{\rm qu},\bar{\omega}_{\rm cl},g_{\rm cl})
=
\sum_{a,b,c=-\infty}^{\infty}\sum_{\alpha,\beta,\gamma=0}^{\infty}W_{\mathrm{path}}^{\nu\alpha\beta\gamma\nu;abc}
\end{equation}
now sums over all path in the photon number and Floquet space;
the weights $W_{\mathrm{path}}^{\nu\alpha\beta\gamma\nu;abc}$ for a path
with photon numbers and Floquet indices $(\nu,0)\to (\gamma,c)\to (\beta,b) \to (\alpha,a) \to (\nu,0)$
are given in the appendix (see Eqs.~\eqref{eq:W_S1_driven}-\eqref{eq:W_DC_driven}).

\subsection{Spin-Photon Hamiltonian approach}\label{sec:SpPh}
Alternative to the derivation presented in the previous section, one could try to start from the spin-photon Hamiltonian~\eqref{eq:spin-phot} (or spin-photon-Floquet Hamiltonian~\eqref{eq:mbot_spin_phot}), from which charge excitations have already been eliminated, and subsequently eliminate the photon excitations.
We will refer to this approach as the ``spin-photon'' approach.
The resulting effective Hamiltonian will be called $H_{\rm eff-SP}$, and the corresponding interaction $\kappa_{\rm SP}$ (using again the parametrization~\eqref{Kparametrization}).
While this is an intuitive procedure, its validity is restricted to certain limits, as will then be discussed in the result section.

\subsubsection{Closed cavity}
Starting from the spin-photon Hamiltonian~\eqref{eq:spin-phot}, we focus on a fixed photon number sector $\nu$ and eliminate all photon number off-diagonal matrix elements.
This corresponds to a more standard second-order perturbation theory, where $H_{\rm eff-SP}$ is given by the first two terms in Eq.~\ref{eq:eff_ham} only, and the off-diagonal matrix elements correspond to the matrix elements $\mathcal{J}$ for $m$-photon emission and absorption.
This gives an interaction $K^{(\nu)}_{\rm SP} = K_0\kappa^{(\nu)}_{\rm SP}(\bar{\omega}_{\rm qu},g_{\rm qu})$, with
\begin{align}
  \kappa^{(\nu)}_{\rm SP} &=
  -2\sum_{n=1}^{\infty}\bra{\nu} \mathcal{J}_{2n} a^{2n} \frac{1}{2n\omega_{\rm qu}} (a^{\dagger})^{2n}\mathcal{J}_{2n}\ket{\nu}\nonumber\\
 &\phantom{=}+2\sum_{n=1}^{\infty}\bra{\nu} (a^{\dagger})^{2n}\mathcal{J}_{2n} \frac{1}{2n\omega_{\rm qu}} \mathcal{J}_{2n}a^{2n}\ket{\nu}.\label{eq:spin-phot-unpert}
\end{align}
The two summands describe virtual photon emission and absorption respectively, where the latter term vanishes for $2n>\nu$.

\subsubsection{Driven cavity}
In the driven case, an initial elimination of charge excitations produces the spin-photon-Floquet matrix elements of Eq.~\eqref{eq:mbot_spin_phot}.
Eliminating both cavity and sideband fluctuation in second order from this effective Hamiltonian (see App.~\ref{sec:drive_spin_phot_int}), we obtain Eq.~\eqref{eq:floq-spin-phot-opp}-\eqref{eq:floq-spin-phot-resolv}.

\section{Results}\label{sec:results}

\subsection{Closed cavity}\label{sec:results_closed}
\begin{figure}
  \centering
  \includegraphics{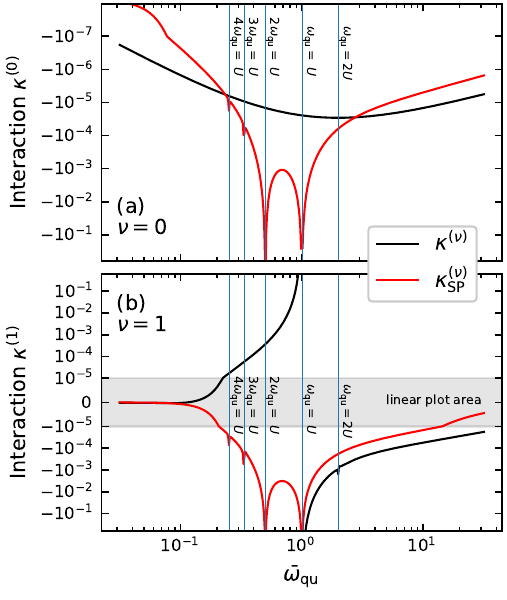}
  \caption{
  Cavity induced singlet-singlet interaction for $g_{\rm qu}=0.1$.
  (a) Empty cavity ($\nu=0$). (b) Cavity with one photon ($\nu=1$).
  The black lines indicate the interaction $\kappa^{{\nu}}$ [Eq.~\eqref{kappa_sum}], while the red lines show the
  interaction $ \kappa^{(\nu)}_{SP}$ obtained from the spin-photon approach [Eq.~\eqref{eq:spin-phot-unpert}].
  In (b) the vertical axis is linear in the shaded area, otherwise it is logarithmic.
}\label{fig:interaction_omega}
\end{figure}
We start by discussing induced interactions for the closed single-mode cavity.
In addition to a general discussion we explicitly compare the full fourth order expansion and the interaction obtained from the simpler spin-photon approach.

The black lines in Fig.~\ref{fig:interaction_omega} show the long-range interaction $\kappa^{(\nu)}$ [Eq.~\eqref{kappa_sum}] as a function of the cavity frequency $\bar{\omega}_{\rm qu}= \omega_{\rm qu}/U$.
(Note that this result, like all other results shown below, is obtained from the analytical expressions of Sec.~\ref{sec:fourthO}, and benchmarked with the numerical approach outlined in Sec.~\ref{sec:lr_overview}.)
The interaction $\kappa^{(0)}$ for an empty cavity, shows a weak dependence on frequency (Fig.~\ref{fig:interaction_omega}(a)) without any singular behavior at the resonances $n\omega_{\rm qu}=U$ ($\bar{\omega}_{\rm qu}=1/n$).
The interaction induced by vacuum fluctuations between two individual dimers is much weaker than the direct spin exchange, unless one reaches the ultra-strong light matter coupling regime ($g_{\rm qu} \gtrsim 1$).
This is understood because the leading order in the long-range interaction is $\mathcal{O}(g_{\rm qu}^4)$.
In Fig.~\ref{fig:interaction_omega}b we also show the result for the isolated cavity with one photon.
In this case, one observes a resonant enhancement of the interaction close to $\bar{\omega}_{\rm qu}=1$.
This resonant behavior comes from an intermediate state with one charge excitation, which is created through virtual absorption of a photon from the cavity, and therefore acquires the energy resolvent $1/(U-\omega_{\rm qu})$.

The red lines in Fig.~\ref{fig:interaction_omega} show the result $\kappa^{(\nu)}_{\rm SP}$ obtained from the spin-photon approach.
One can see that for the undriven cavity with few photons, the spin-photon approach generally gives an incorrect result:
For $\bar{\omega}_{\rm qu}\gg 1$ we find $\kappa^{(\nu)}_{\rm SP} \cong \kappa^{(\nu)}/4$ (see discussion below for the factor $1/4$), while at smaller frequencies the prediction based on the spin-photon Hamiltonian is also qualitatively wrong.
In particular, the spin-photon approach predicts a resonant enhancement of the interaction at integer fractions $\bar{\omega}_{\rm qu}= 1/n$ even for the empty cavity.
These divergences arise from the divergence of the photon number off-diagonal matrix elements $\mathcal{J}_{2n}$ in the spin-photon Hamiltonian~\eqref{eq:spin-phot}, or, in more physical terms, the resonant enhancement of the optical non-linearity.
In contrast, it is clear that there should be no resonant enhancement of the induced interaction for the empty cavity, because all intermediate states which contribute to the correlated super-exchange are gapped from the ground state.

While the deviation between the approaches is not too surprising from a formal standpoint, it is nevertheless of physical importance:
In order to understand the cavity-mediated interactions, one in general cannot use a phenomenological approach that would use the spin-photon Hamiltonian with matrix elements that are obtained from nonlinear optical measurements (see comments at the end of Sec.~\ref{sec:sp_phot}).

\subsection*{Understanding the differences between the approaches}

We can understand the deviation between the spin-photon approach and the exact result, if we take a closer look at how they are derived.
Writing out the transformations, the two effective Hamiltonians are given by
\begin{align}
  H_{\rm eff}   &= P^{c,U}_0 e^{S_{c,U}} H e^{-S_{c,U}} P^{c,U}_0,\\
  H_{\rm eff-SP} &= P^c_0 e^{S_c} P^U_0 e^{S_U} H e^{-S_U} P_0^U e^{-S_c} P^c_0,\label{HSP-schematic}
\end{align}
where $H_{\rm eff}$ is the effective Hamiltonian obtained from simultaneous elimination of charge ($U$) and cavity ($c$) excitations, while $H_{\rm eff-SP}$ is obtained from the spin-photon Hamiltonians, i.e., from the successive elimination of both degrees of freedom (Sec.~\ref{sec:SpPh}).
Here $S_{c,U}$, $S_c$, $S_U$ are the generators of the unitary transformations and $P^{c,U}_0$, $P^c_0$, $P^U_0$ the corresponding projectors to the target subspace.
For the spin-photon approach, one first constructs the spin-photon Hamiltonian $H_{\rm SP}=P^U_0 e^{S_U} H e^{-S_U} P_0^U$ in the doublon free subspace, and then removes the photons.
One deviation which can be directly read of from the definition is caused by the position of the projector $P_0^U$:
The first unitary transformation applied to $H$ in Eq.~\eqref{HSP-schematic} ($e^{S_U}$) removes the leading-order coupling between the doublon free subspace and the rest of the Hilbert space.
Subsequently projecting to the doublon free subspace removes the contribution of intermediate states with more than one double occupation.
The projection to the doublon free subspace therefore cuts out all contributions of the DC-paths from the perturbation theory.
Furthermore, since the interaction is only mediated by the photon number offdiagonal parts of $H_{\rm SP}$, the contributions from the S1-paths (or rather the imbalance between the DC- and S1-path) are not captured either.
This is easiest seen in Eq.~\eqref{eq:spin-phot-unpert}, where the series expansion does not provide a contribution for $n=0$.
\begin{figure}
  \centering
  \includegraphics{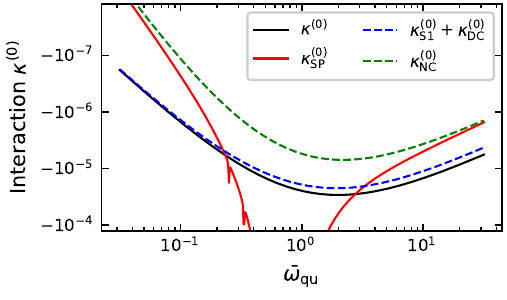}
  \caption{
    Different contributions (dashed) to the total interaction $\kappa^{(0)}$ (black) in the ground state of the quantum field.
    For $\omega_{\rm qu} \gg U$, the spin-photon based interaction $\kappa_{\rm SP}^{(0)}$ (red) converges to $\kappa_{\rm NC}^{(0)}=\kappa^{(0)}/4$.}\label{fig:interaction_omega_add}
\end{figure}

To support this discussion, it is illustrative to individually compare the contributions $\kappa^{(\nu)}_{\rm NC}$ and $\kappa^{(\nu)}_{\rm S1} + \kappa^{(\nu)}_{\rm DC}$\footnote{Since the Schrieffer-Wolff transformation is not a linked cluster expansion, the S1- and DC-paths will individually give contributions to an interaction even in the uncoupled case. Only when considering both of them together, the linked cluster property of the effective Hamiltonian is restored. To avoid comparing the proper interaction in the coupled case with artifacts of unlinked clusters, we will only consider the sum of all S1- and DC-paths.} to the interaction $\kappa^{(\nu)}_{\rm SP}$ obtained from the spin-photon approach (see Fig.~\ref{fig:interaction_omega_add}).
In the far-off resonant limit $\omega_{\rm qu}\gg U$ one finds that $\kappa^{(\nu)}_{\rm SP}$ approaches the contribution $\kappa^{(\nu)}_{\rm NC}$.
This not surprising, because the NC-paths have the same structure as that imposed by the spin-photon approach.
Moreover, in this limit the NC-paths contribute precisely $1/4$ of the interaction. 
It is illustrating to confirm these observations from the analytical expressions to leading order in $g_{\rm qu}$:
Expanding Eqs.~\eqref{eq:s1}-\eqref{eq:DC} (with~\eqref{eq:s1path}-\eqref{eq:DCpath}) in the light-matter coupling $g_{\rm qu}$
(see App.~\ref{AppD2}), we find
\begin{align}
\kappa^{(0)} &= -2 g_{\rm qu}^4 \frac{\bar{\omega}_{\rm qu}^2}{(1+\bar{\omega}_{\rm qu})^3}
+\mathcal{O}(g_{\rm qu}^6),
\\
\kappa_{\rm NC}^{(0)} &= -2g_{\rm qu}^4\frac{\bar{\omega}_{\rm qu}^3}{(1+\bar{\omega}_{\rm qu})^2(1+2\bar{\omega}_{\rm qu})^2}
+\mathcal{O}(g_{\rm qu}^6),
\end{align}
while the spin-photon approach gives
\begin{align}
  \kappa^{(0)}_{\rm SP}= -2\frac{|\bra{0}\mathcal{J}_2\ket{0}|^2}{\bar \omega_{\rm qu}} + \mathcal{O}(g_{\rm qu}^6),
\end{align}
with $\mathcal{J}_2$ given by Eq.~\eqref{eq:phot_offdiag_pert}.
For $\bar\omega_{\rm qu} \gg 1$, one can now see that \mbox{$\kappa^{(0)}_{\rm SP}=\kappa_{\rm NC}^{(0)}=  \kappa^{(0)}/4$}.

The divergences of $\kappa_{\rm SP}$ at the resonances have a different origin.
Since the first elimination in the derivation of $H_{\rm eff-SP}$ is a multi-block orthogonalization scheme (each cavity occupation number defines a block), it relies on proper energy gaps between all of these blocks in the unperturbed Hamiltonian.
Close to resonance between charge and photon excitations, the condition $|n\omega_{\rm qu} - U|\gg t_0$ necessary for a convergent series expansion is no longer fulfilled for some integer $n$, which leads to artificial resonances in the spin-photon based perturbation theory at $\bar{\omega}_{\rm qu} = 1/n$.
To leading order $g_{\rm qu}^4$, we find these resonances at $\bar{\omega}_{\rm qu}= 1$ and  $1/2$, where $\mathcal{J}_2$ [Eq.~\eqref{eq:phot_offdiag_pert}] diverges.
Higher orders produce additional divergences at higher ratios.
The fourth-order approach, in contrast, is a two-block scheme.
It therefore only requires the target sector to be sufficiently gapped to the rest of the Hilbert space, which is always fulfilled for $t_0\ll \omega_{\rm qu},U$.

The spin-photon approach is therefore expected to properly describe the interactions only if the result is dominated by the NC-paths and they themselves are properly contained.
We find regimes where both holds in the driven setting, when the classical drive is near resonant to the cavity.

\subsection{Results: Driven cavity}\label{sec:results_driven}
\begin{figure}
  \centering
  \includegraphics{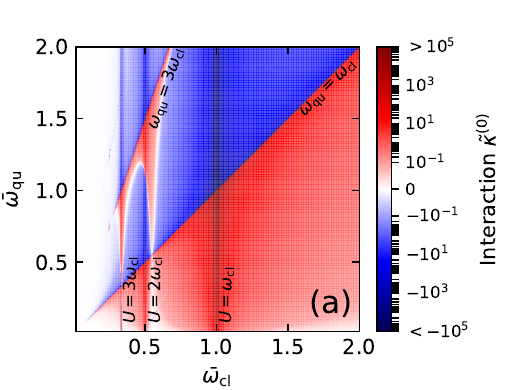}\\
  \centering
  \includegraphics{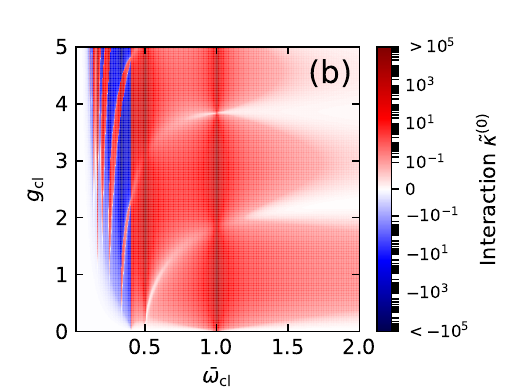}
\caption{
(a) Induced long-range interaction $\tilde{\kappa}$ [Eq.~\eqref{Ktilde}] as function of the classical and the cavity frequency for \mbox{$g_{\rm cl}=0.7$};
(b) Interaction as function of the classical frequency $\omega_{\rm cl}$ and driving strength $g_{\rm qu}$, for $\bar{\omega}_{\rm qu}=0.4$.
For both plots the logarithmic scale of the colomap is interrupted by a linear scale between $-10^{-1}$ to $10^{-1}$.}\label{fig:driven}
\end{figure}

In this section, we proceed to the discussion of the driven cavity.
Probably the experimentally far most relevant setting is a situation where the coupling to the quantum field is weak, and an external drive is used to boost and control the interaction.
For example, the cavity field can correspond to surface plasmon mode, and a laser is used to ``activate'' the exchange of virtual plasmon, which then mediate the interaction (see also discussion in Sec.~\ref{sec:real_systems}).
We therefore restrict the discussion of the driven cavity to results which are leading order in the coupling $g_{\rm qu}$, but of arbitrary order in the laser amplitude $g_{\rm cl}$.
Moreover, all results for the driven case will be restricted to $\nu=0$, and we therefore omit the index $\nu$ in the following.
The interaction~\eqref{Kparametrization} is therefore written as
\begin{align}
\kappa(\bar{\omega}_{\rm qu},\bar{\omega}_{\rm cl},g_{\rm cl},g_{\rm qu}) \equiv
  g_{\rm qu}^2 \tilde{\kappa}(\bar{\omega}_{\rm qu},\bar{\omega}_{\rm cl},g_{\rm cl}) + \mathcal{O}(g_{\rm qu}^4),
\label{Ktilde}
\end{align}
and we will analyze the result $\tilde{\kappa}$, and the corresponding expression $\tilde{\kappa}_{\rm SP}$ from the spin-photon approach.

Fig.~\ref{fig:driven}(a) shows the interaction $\tilde{\kappa}$ for fixed driving strength $g_{\rm cl}$ as function of $\omega_{\rm qu}$ and $\omega_{\rm cl}$.
As before, the data are obtained with the full fourth-order approach and benchmarked against exact diagonalization.
From the colormap, we can see that it is possible to enhance the long-range interactions using two different types of near-resonant driving:
One option is driving the Mott gap resonantly, taking
\begin{equation}
|\bar{\Delta}_{U}| \equiv |\bar{\omega}_{\rm cl}-1| \ll 1.
\end{equation}
This amounts to the vertical strip around $\bar{\omega}_{\rm cl}=1$ in the figure, where the interaction diverges like $\bar \Delta_{U}^{-2}$.
Alternatively, we resonantly drive the quantum field as $ |\omega_{\rm qu} - \omega_{\rm cl}| \ll t_0$, i.e.,
\begin{equation}
|\bar{\Delta}_{\rm qu}| \equiv |\bar{\omega}_{\rm qu} - \bar{\omega}_{\rm cl}| \ll 1.
\end{equation}
This amounts to the diagonal line $\bar{\omega}_{\rm qu}=\bar{\omega}_{\rm cl}$ in the color plot, where the interaction diverges like $\bar{\Delta}_{\rm qu}^{-1}$.
At sufficiently large driving strengths $g_{\rm cl}$, there are also
singularities at the multi-photon resonances, such as $n\omega_{\rm cl} =U$, or $n\omega_{\rm cl} =\omega_{\rm qu}$.
Figure~\ref{fig:driven}(b) shows the dependence  of the interaction on the driving strength $g_{\rm cl}$.
One observes a rich behavior with many zeros and sign changes.
These are associated with the zeros of the Bessel functions in the Floquet Hubbard model~\eqref{eq:mat_floq_cav}, corresponding to dynamical localization of the electrons.
\begin{figure}
   \includegraphics{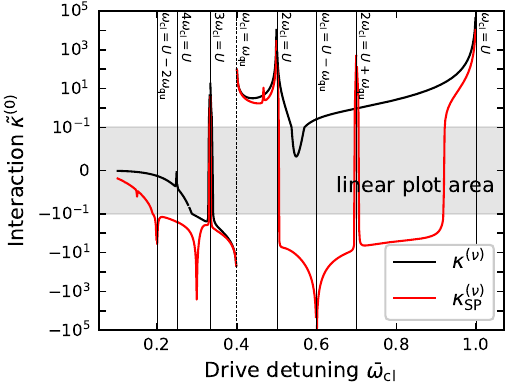}
\caption{
Horizontal cross-section of the data in Fig.~\ref{fig:driven}(a) at $\bar{\omega}_{\rm qu}=0.4$, compared to the interaction $\tilde{\kappa}_{\rm SP}$;
As with the undriven case, the spin-photon based approach has additional resonances, but matches the proper approach at $\bar \Delta_{\rm qu} \ll 1$ (dashed line).
In the grey-shaded area, the scale of the plot is again linear.}\label{fig:driven-cut1}
\end{figure}

To access the validity of the spin-photon approach, we compare the interactions $\tilde{\kappa}$ and $\tilde{\kappa}_{\rm SP}$ along a cut of constant $\bar{\omega}_{\rm qu}$ in Fig.~\ref{fig:driven}(a), see Fig.~\ref{fig:driven-cut1}.
One finds that the leading resonance at $\omega_{\rm qu}=\omega_{\rm cl}$ is captured by both approaches (dashed line), while away from this resonance the two approaches deviate.
The spin-photon approach again features additional divergences, and it has an opposite sign in some regimes.

The fact that the leading resonance in  $\bar{\Delta}_{\rm qu}$ is
captured by the spin-photon approach can be explained as follows:
For $\bar{\Delta}_{\rm qu}\ll1 $ the interaction is dominated by an intermediate state without electronic excitations, but only the exchange of a photon from the drive to the cavity.
This corresponds to an exchange path such as shown in Fig.~\ref{fig:resonant}, in which the processes on the two dimers i.e., Raman-type processes leading to a spin flip upon exchange of a photon between laser and cavity, can be understood as successive.
This successive picture is precisely contained in the spin-photon approach.

Mathematically, expanding the long-range interaction in the detuning
$\bar{\Delta}_{\rm qu}$~\footnote{Since the interaction diverges at this resonance, the leading order will be $\bar{\Delta}_{\rm qu}^{-1}$.}, we find that the leading contribution $\propto 1 / \Delta_{\rm qu}$ from the NC-paths and the driven spin-photon Hamiltonian agree (see App.~\ref{sec:drive_spin_phot_int}, Eq.~\ref{eq:resolvent_detuning}).
In this region the interaction is therefore properly captured and the phenomenological ansatz valid.

If we instead choose $\bar{\Delta}_U$ small, i.e., near resonant driving of charge excitations, the scalings of the three paths become $H_{S1} \propto \Delta_U^{-3}$, $H_{NC} \propto \Delta_U^{-2}$ and $H_{DC} \propto \Delta_U^{-3}$, such that the resonant S1/DC-paths dominate and the leading order is not correctly captured by the spin-photon approach.

\section{Discussion for realistic parameters}\label{sec:real_systems}

\subsection{Single mode cavity setting}

In this section, we illustrate the previous results for a realistic set of parameters.
For the matter we assume a lattice constant $d=1\,$nm, Hubbard interaction $U=0.8\,$eV, and hopping $t_0=50\,$meV.
These parameters are close to the organic Mott insulator ET-F$_2$TCNQ~\cite{Hasegawa1997,Mitrano2014}\footnote{Note that in ET-F$_2$TCNQ one should also consider a nearest neighbor interaction $V$. The parameter $U$ in our formalism measures the energy of a doublon-hole excitation on a dimer, and is therefore given by $U=U_{\rm loc}-V$, with a local Hubbard $U_{\rm loc}$ and a nearest neighbor interaction. Both $t_0$ and $V$ can be tuned by  pressure over some range~\cite{Mitrano2014}.}, but can be taken in general as representative for a good Mott insulator with a small ratio $t_0/U$.

The cavity is modelled by a single mode resonator, where the electric field is confined in a volume $L^3$ with a homogeneous mode function.
This can be taken as an ideal description of a split-ring resonator~\cite{Faist2014}.
The assumption of a cubic volume and a homogeneous mode function is of course rather simplistic, but it gives the correct order of magnitude for the light-matter coupling.
Realistic settings allow for resonance frequencies $f_{\rm qu} = \omega_{\rm qu}/2\pi$ in the THz regime, and a $\mu$m-sized cavity, which corresponds to a large compression of the mode volume ($L^3$) below the free space value $\lambda_{\rm qu}^3= ( c/f_{\rm qu})^3$.
We will exemplarily consider a cavity frequency $f_{\rm q}=\omega_{\rm q}/2\pi=6\,$THz.
In this case $\hbar\omega_{\rm qu}$ is sufficiently small compared to the charge gap $U$ ($\bar{\omega}_{\rm qu}\approx 0.0341$), such that electronic excitations due to (multi)-photon absorption are strongly suppressed for driving with an external laser at a frequency $\omega_{\rm cl}$ close to $\omega_{\rm qu}$.
For further illustration, we will also consider larger frequencies (such as $f_{\rm q}=\omega_{\rm q}/2\pi=60\,$THz, $\bar \omega_{\rm qu}\approx 0.341$) for which cavities may be more difficult to design, but which is still sufficiently detuned from the charge gap.
 
For a single mode with electric field confined in the volume $L^3$, standard quantization gives the vaccum field strength [c.f.~Eq.~\eqref{Aqu}]  $A_{\rm qu} = [\hbar/(2\epsilon_0\omega_{\rm qu} L^3)] ^{1/2}$.
(We restore factors $\hbar$ in this section.)
With the Peierls phase~\eqref{peierls}, the dimensionless coupling $g_{\rm qu} = A_0 d q/\hbar$ becomes
\begin{align}
g_{\rm qu} 
=
\sqrt{\frac{e^2d^2}{2\epsilon_0 L^3 \hbar\omega_{\rm qu}}}
\approx 
\frac{46.8}{ \sqrt{f_{\rm qu}[{\rm THz}]}} \sqrt{\frac{d^3}{L^3}}.
\label{gqnumebrs}
\end{align}

For this setting, we will now compute the cavity-induced long-range interaction, and compare it to the other
relevant scale, the short-range exchange $J$. For this, a few comments are in order: 

(i) In the (driven) cavity, also the direct exchange $J$ will be modified with respect to the free space value $J_0=2t_0^2/U$. However, because we are mainly interested in quantifying the strength of the induced long-range interactions, we compare the long-range interaction to the same scale ($J_0$) for all parameters.

(ii) The effect of the long-range interaction on the material depends, in addition to the strength of the interaction, on the geometry.
For example, one can imagine a 2D geometry, where the direction of the cavity polarization implies that long-range interactions are induced only along one direction.
An interaction $K<0$  would therefore favor the bonds along that direction to be in a singlet state, in competition with the isotropic Heisenberg exchange.
In this paper we focus on the strength of the induced interactions (and how to compute them), while the discussion of possible phase transitions due to such interactions is left for future work.
We will therefore evaluate a relevant overall scale of the interaction, defined as follows:

(iii) For the single-mode cavity, the induced interaction is an all-to-all interaction.
Hence, the short range exchange should not be compared to the interaction $K$ between individual dimers, but to the mean-field interaction $K_{\rm mf}=NK$ of one given dimer with all ($N$) others.
For simplicity, we assume that the cavity is filled with the material, so that $N=(L/d)^3$, and analyze the ratio
\begin{align}
\frac{K_{\rm mf}}{J_0}
=
\frac{8K_0}{J_0} \frac{L^3}{d^3} \kappa^{(\nu)}(g_{\rm cl},g_{\rm qu},\bar{\omega}_{\rm cl},\bar{\omega}_{\rm q}),
\end{align}
where $\kappa$ is computed as in the previous sections. For $\frac{K_{\rm mf}}{J_0}\gtrsim 1$, one can expect the cavity-induced long-range interactions to become a relevant or even dominant correction to the short-range Heisenberg exchange.

\subsection{Undriven cavity}

\begin{figure}
 \begin{center}
  \includegraphics{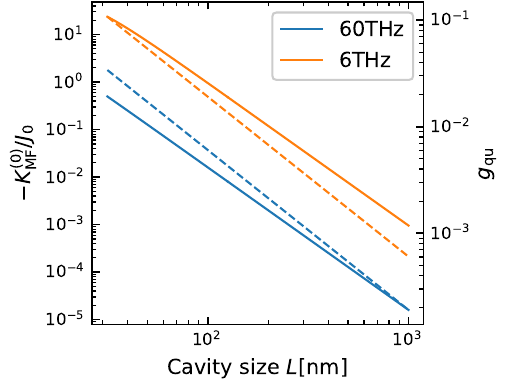}
  \end{center}
  \caption{
  Ratio $K_{\rm mf}/J_0$ (solid curves, left axis) for the empty cavity, where the size of the cavity is changed in order to control the light-matter coupling $g_{\rm qu}$ (dashed curves, right axis). Parameters are $U=0.8\,$eV, $t_0=50\,$meV, $d=1\,$nm, and cavity frequencies $f_{\rm qu}=6\,$THz ($\bar \omega_{\rm qu}\approx 0.0341$) and $f_{\rm qu}=60\,$THz ($\bar \omega_{\rm qu}\approx 0.341$) as indicated.
  }\label{fig:real-quantum}
\end{figure}
Figure~\ref{fig:real-quantum} shows the ratio $\frac{K_{\rm mf}}{J_0}$ for the undriven cavity.
One can see that a strong long-range interaction can be reached only for relatively small cavities, while the effect of the interaction vanishes for large $L$.
This can be understood as follows:
For large $L$, the coupling decreases with increasing mode volume like $g_{\rm qu}\sim L^{-3/2}$, and the induced interaction can therefore eventually be approximated by the leading order in $g_{\rm qu}$, i.e., $K \sim g_{\rm qu}^4\sim L^{-6}$.
In the thermodynamic limit (being defined as $L\to\infty$, $g_{\rm qu}\sim L^{-3/2}$, and  $N\sim L^3$), interactions which are induced by the vacuum fluctuations of a single mode therefore scale like $K_{\rm mf}=KN \sim L^{-3}$ and  become irrelevant.
The scaling should hold similarly for other interactions which are induced by nonlinear processes.
This finding is in line with general arguments which imply that the change of of the energy of an extended material ($\sim N$ atoms) due to the coupling  to a {\em single} cavity mode is $\mathcal{O}(N^0)$ (sub-extensive) in the thermodynamic limit, and therefore irrelevant for the static properties of the material~\cite{Pilar2020,Lenk2022b,Bernardis2018}.

On the other hand, the quantitative analysis in Fig.~\ref{fig:real-quantum} shows that a ratio of $K_{\rm mf}\approx J_0$, where the long-range interactions become comparable or even dominant over the short-range interactions, can be obtained already for cavities which are still large enough to host a quasi macroscopic number of atoms (e.g., for $f_{\rm qu}=6\,$THz, $K_{\rm mf}/J_0 \approx 1$ for $L\approx100\,$nm, corresponding to $N\sim 10^6$ unit cells).
Hence, collective many-body effects due to vacuum-induced long-range interactions may indeed be accessible under realistic conditions.
At the same time, one should note that in this parameter regime the single particle coupling $g_{\rm qu}$ is still considerably smaller than one.
Because the cavity effect on the short range Heisenberg exchange interactions is directly controlled by $g_{\rm qu}$ without an additional factor $N$ (see, e.g.,~Eq.~\eqref{eq:phot_diag_pert}), long-range interactions constitute the main effect of the cavity on the material in this regime.
For even smaller cavities, long-range interactions dominate over the short-range Heisenberg exchange (until the system is too small to be considered as macroscopic, such as in a nano-plasmonic cavity).
This shows that in general long-range interactions should be kept in mind whenever single-mode cavity settings are proposed to engineer spin or other type of exchange interactions~\cite{Li2022,Bostroem2022,Kiffner2019,Sentef2020}.

As a side remark, note that the vanishing of the interaction in the thermodynamic limit naturally implies the absence of any phase transition induced by the empty single-mode cavity, at least in the strict mathematical sense.
In particular, this applies to a hypothetical phase transition where due to the induced long-range interaction a macroscopic number of dimers would ``condense''.
This condensation  would induce a macroscopic squeezing of the cavity mode ($\langle a^2\rangle\sim N$, $\langle a\rangle=0$), in analogy to the equilibrium superradiant phase with $\langle a\rangle\sim N$~\cite{Hepp1973,Latini2021,Mazza2019}.
The superradiant phase is absent for the single mode case when linear terms $\propto A$ and quadratic (diamagnetic) terms $\propto A^2$ of the light matter interaction are treated consistently~\cite{Andolina2019,Andolina2020}, and a similar argument should hold for the hypothetical macroscopic squeezing transition with respect to the nonlinear light-matter interactions contained in the Peierls phase.

\subsection{Driven cavity}
\begin{figure}
 \begin{center}
  \includegraphics{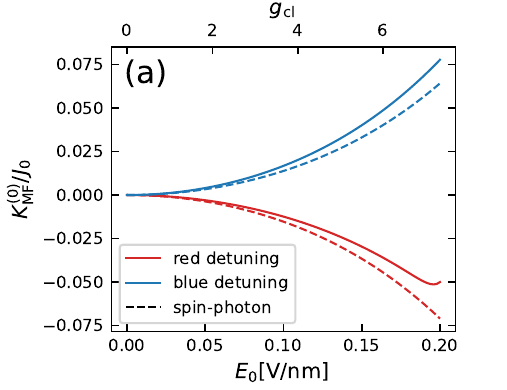}
  \includegraphics{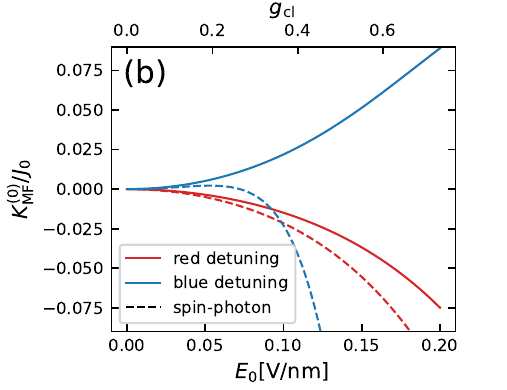}
  \includegraphics{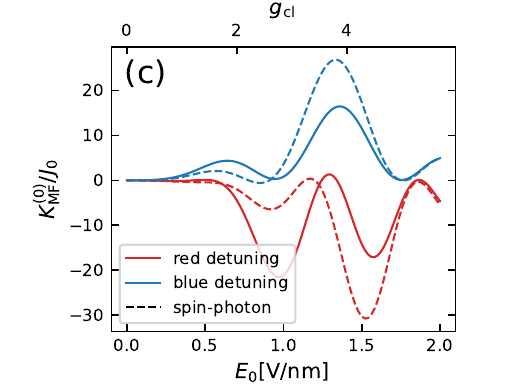}
 \end{center}
 \caption{
Ratio $K_{\rm mf}/J_0$ for the driven cavity in the large $L$ limit
where red and blue detuning denote a relative detuning $\mp 10\%$ of the driving frequency with respect to the cavity frequency.
The three cases displayed are (a) $f_{\rm qu} = 6\,$THz, (b) $f_{\rm qu} = 60\,$THz, (c) $f_{\rm qu} = 77.4\,$THz. The latter amounts to the parameters of Fig.~\ref{fig:driven}(b) and can be understood as vertical cuts left and right of the resonance $\omega_{\rm cl}=\omega_{\rm qu}$.
In all plots, the solid lines indicate the interaction as obtained by the fourth-order approach, the dashed lines are obtained by the spin-photon approach.
  }\label{spliringdriven}
\end{figure}

We now proceed to the driven case.
As explained in Sec.~\ref{sec:results_driven}, a near-resonant laser drive $\omega_{\rm cl}\approx\omega_{\rm q}$ is favourable condition to enhance the interaction without heating the material.
However, because real cavities in condensed matter setting can have relatively low quality factors $Q$ (such as a $Q\approx10$~\cite{Faist2014}), the detuning cannot become too  small; otherwise the drive would populate the cavity.
We therefore analyze a laser frequency with $\pm10\%$ detuning, e.g., $f_{\rm cl}=0.9\cdot\omega_{\rm qu}/2\pi=5.4\,$THz for the red detuned case with respect to $f_{\rm qu}=6\,$THz.
The laser coupling strength is $g_{\rm cl}=deE_0/(\hbar\omega_{\rm cl})$, with the electric field amplitude $E_0$.
For $d=1\,$nm, $g_{\rm cl}$ can be written as
\begin{align}
g_{\rm cl}
\approx
242\,
\frac{E_0 [{\rm Vnm}^{-1}]}{f_{\rm cl}[\rm THz]}.
\end{align}
Hence, non-perturbative couplings $g_{\rm cl}\gtrsim 1$ can be reached with field strength of the order $0.1\,$V/nm.
This should be experimentally accessible, in particular taking into account near-field enhancement effects.

Although the formalism provided in the main text also applies when both $g_{\rm cl}$ and  $g_{\rm qu}$ are strong, the driven case is most interesting in the regime where the interaction induced by vaccum fluctuations is weak.
This is the case for large $L$, where the vacuum-induced interaction scales like $K_{\rm mf} \sim N g_{\rm qu}^4\sim L^{-3}$.
In contrast, in the driven case the interaction between individual dimers scales like $g_{\rm qu}^2\sim L^{-3}$ [c.f.~Eq.~\eqref{Ktilde}], so that $K_{\rm mf} \sim L^3 g_{\rm qu}^2$ remains finite in the thermodynamic limit.
In Fig.~\ref{spliringdriven} we therefore show the large $L$ limit of the  induced interaction, $K_{\rm mf}/J_0=K_0/J_0 (Ng_{\rm qu}^2)  \tilde{\kappa}(\bar{\omega}_{\rm qu},\bar{\omega}_{\rm cl},g_{\rm cl}) $, where the factor $Ng_{\rm qu}^2\approx (46.8)^2/f_{\rm qu}[{\rm THz}]$ is independent of $L$ [Eq.~\eqref{gqnumebrs}].

Figure~\ref{spliringdriven} shows the interaction ratio as a function of the laser driving $g_{\rm cl}$ (or $E_0$), for several values of $f_{\rm qu}$, and fixed detunings $f_{\rm cl}= (1\pm0.1)f_{\rm qu}$.
One finds that with sufficiently large drivings, interactions ratios $K_{\rm mf}/J_0$ of order $1$ can be induced, even for drivings which are sufficiently off-resonant for low Q factors.
At the same time, the comparison with the result obtained from the spin-photon Hamiltonian (dashed line) can deviate qualitatively from the full prediction.
This clearly shows that in general, to discuss laser-induced long-range spin interactions in solids, a correct treatment of the off-resonant terms, as done by the full fourth-order theory provided in this work, is necessary.
This should hold similarly for other interactions which are induced by nonlinear processes.

\section{Conclusion and outlook}
\label{sec:concl}

In summary, we have discussed  cavity-mediated long-range interactions within a Mott insulator, as described by the Hubbard model.
These interactions correspond to correlated spin flips at distant sites, and originate from the nonlinear interplay between spins and photons, such as Raman scattering and multi-photon absorption/emission.
We have explored  these interactions in two  scenarios: (i) In the undriven cavity, where interactions emerge through the exchange of virtual photons, and (ii),  in a laser-driven cavity, which opens the potential for Floquet engineering of long-range interactions.

In the derivation of these long-range interactions, a simple approach would be to start from an effective Hamiltonian which describes the nonlinear interaction of photons and spins.
At first sight, this ``spin-photon'' approach  seems intuitive; e.g., vacuum mediated interactions would be obtained to second order in the  effective light-matter interaction, with virtual two-photon emission and absorption at two different lattice sites.
Moreover, it  would allow for a simple phenomenological determination of the relevant matrix parameters by optical measurements in free space (such as Raman scattering).
However, we show that the spin-photon approach has its limitations.
It can be used only within narrow parameter regimes, particularly when the laser is in close resonance with the cavity resonance.
In typical condensed matter systems, these resonance conditions may not be easy to meet, e.g., when cavities of low $Q$ factor are used and when the material itself has broad absorption bands.
Moreover, the spin-photon approach gives a qualitatively wrong prediction of the vacuum-mediated interactions.
Instead, we have provided a comprehensive derivation of the interactions, starting from the underlying electronic model (a Hubbard model), using a fourth-order perturbation theory in the parameter $t_0/U$.

We have evaluated these interactions for a single mode resonator, such as a split-ring cavity.
While one can see that in this case the effect of long-range interactions eventually becomes negligible in the thermodynamic limit (because the light-matter coupling decreases like $\sim 1/\sqrt{V}$ with the mode volume), one finds that the light-induced interactions can remain highly relevant and even dominate over the short-range Heisenberg interactions up to cavity sizes which still host an almost macroscopic ensemble of atoms.
This motivates future studies of the nonlinear response of correlated electron systems in mesoscopic settings such as small split-ring resonators.

This work naturally extends in several directions:
(i) The long-range interactions are rather unconventional, as they do not correspond to long-range Heisenberg interactions, but to correlated spin flips at distant sites (i.e., four-spin processes).
It will be interesting to study whether these interactions can lead to exotic magnetic orders  by competing with the short range Heisenberg interactions.
(ii) The analysis is not limited to spin systems but can encompass all degrees of freedom that nonlinearly couple to light.
Systems with orbital order, which often show frustration of the short-range interaction, represent an interesting material class in this regards.
(iii) The derivation can naturally extend to a multi-mode case, where a single cavity mode is replaced by a continuum of modes, such as coplanar waveguides, Fabry-Perot cavities, or surface plasmons~\cite{Schlawin2019, Chiocchetta2021, Li2022, Lenk2022, Ashida2022}.
This is in particular relevant for the laser-driven scheme: A photon can be scattered from the laser to the dispersive cavity mode and back into the laser, and thus mediate an interaction $K(\bm r,\bm r')$ between  scattering centers at sites $\bm r$ and $\bm r'$, where the dependence on distance $\bm r-\bm r'$ is set by the dispersion of the cavity mode.
As long as  the interaction to the quantum mode is treated to leading order in the coupling (c.f.~Eq.~\ref{Ktilde}), the results of our work should be generalizable more or less by replacing  $g_{\rm qu}^2$ by the space dependent couplings $g_{\bm q}(\bm r)g_{\bm q}(\bm r')^*$ for a mode with wave vector $\bm q$,  and then summing over all modes $\bm q$.

\section*{Acknowledgements}
P.F. and M.E. were funded by the ERC Starting Grant No. 716648;
K.P.S., P.F. and M.E. by the Deutsche Forschungsgemeinschaft (DFG, German Research Foundation), Project-ID No. 429529648, TRR 306 QuCoLiMa (“Quantum Cooperativity of Light and Matter”);
J.L. thanks the funding from the European Union’s Horizon 2020 research and innovation programme under the Marie Sklodowska-Curie grant agreement No. 884104.

\onecolumngrid{}

\appendix

\section{Explicit for of the spin-photon Hamiltonian }\label{sec:OpForm}

In~\cite{Sentef2020}, the second-order time-dependent Schrieffer-Wolff transformation for the cavity coupled Hubbard model yields the effective spin-photon Hamiltonian:
\begin{align}
  H &= J_{\mathrm{ex}}\sum_{\langle i,j \rangle}\left( \vec{S}_i\vec{S}_j - \frac{1}{4} \right) \mathcal{J}[a^{\dagger},a] + \omega_{\rm qu} a^{\dagger}a\\
 \mathcal{J}[a,a^{\dagger}] &= \mathcal{J}_0[a^{\dagger},a] + \sum_{m=1}^{\infty}\left( \left( a^{\dagger} \right)^{2m}\mathcal{J}_{2m}[a,a^{\dagger}] + \mathrm{h.c.}\right)\\
  \mathcal{J}_{2m}[a,a^{\dagger}] &= \sum_{c=0}^{\infty}g_{\rm qu}^{2c+2m} \left( a^{\dagger} \right)^c a^c \mathcal{L}_{c,m}(g_{\rm qu},\bar{\omega})\\
  \mathcal{L}_{c,m}(\bar{\omega},g_{\rm qu}) &= \frac{1}{2(2c + 2m)!c!}\sum_{p=0}^{2(c+m)}(-1)^p {2(c+m)\choose p}(L_{p-c}(\bar{\omega},g_{\rm qu}) + L_{p-c-2m}(\bar{\omega},g_{\rm qu}))\\
  L_p(\bar{\omega},g_{\rm qu}) &= e^{-g_{\rm qu}^2}\sum_{r=0}^{\infty}\frac{g_{\rm qu}^{2r}}{r!}\frac{1}{1+(r+p)\bar{\omega}}
\end{align}
with $J_{\mathrm{ex}}=4t_0^2/U$ and $\bar{\omega}=\omega_{\rm qu}/U$.

\section{Derivation of the driven spin-photon Hamiltonian}\label{eq:app_driv_spin_phot}

In this section, we perform the second order Schrieffer Wolff transformation to eliminate double occupancies from the driven cavity coupled Hamiltonian~\eqref{eq:mat_floq_cav}, following the general scheme outlined in Sec.~\ref{SecSWT}.
The low energy target space $\mathcal{P}_0$ is therefore the doublon free sector, and the target space projector $P_0$ is given by
\begin{equation}
P_0 = \prod_i \left( 1 - n_{i\uparrow}n_{i\downarrow} \right).
\end{equation}
To leading order, the generator~\eqref{eqs1generat} of the Schrieffer Wolff transformation therefore becomes
\begin{align}
  P_0S_1P_1 &= \sum_{i\in \mathcal{P}_0,j\in \mathcal{P}_1} \frac{\ket{i}\bra{i}V\ket{j}\bra{j}}{E_i - E_j}
          =
          \sum_{m,l=-\infty}^{\infty}\sum_{\mu,\lambda=0}^{\infty}
           \frac{P_0V^{ml}_{\mu\lambda}P_1 \ket{m,\mu}\bra{l,\lambda}}{0+m\omega_{\rm cl}+\mu\omega_{\rm qu}-(U+l\omega_{\rm cl}+\lambda\omega_{\rm qu})},\label{ggegheeqqq}
\end{align}
where $V$ is the hopping part $(\sim t_0)$ of the Hamiltonian.
In the second equality, we represent the operator in the Floquet/photon basis $\ket{m,\mu}$, and $V^{mn}_{\mu\nu}$ is the corresponding electronic part of the operator elements, as implicitly defined in Eq.~\eqref{eq:mat_floq_cav}, i.e.,
\begin{align}
V^{mn}_{\mu\nu}=
- t_0\,\im^{|m-n|+|\mu-\nu|}J_{|m-n|}(g_{\rm cl})j_{\mu,\nu}(g_{\rm qu})
\sum_{\left\langle i,j \right\rangle \sigma}\chi_{ij}^{m-n}\xi_{ij}^{\mu-\nu}\,c^{\dagger}_{i\sigma}c^{\phantom{\dag}}_{j\sigma}.\label{Vqwer}
\end{align}
Analogous to Eq.~\eqref{ggegheeqqq}, we have
\begin{align}          
  P_1S_1P_0 &= \sum_{l,n=-\infty}^{\infty}\sum_{\lambda,\nu=0}^{\infty} \frac{P_1V^{ln}_{\lambda\nu}P_0 \ket{l,\lambda}\bra{n,\nu}}{U+l\omega_{\rm cl}+\lambda\omega_{\rm qu}-(n\omega_{\rm cl}+\nu\omega_{\rm qu})}
\end{align}
The second order the Schrieffer-Wolff transformation yields the effective Hamiltonian given by the leading two terms in Eq.~\eqref{longheff}.
Taking its matrix elements with respect to the cavity occupation and Floquet sidebands,
\begin{align}
  \tilde{H}^{mn}_{\mu\nu} &= \bra{\mu,m} P_0\left( H_0 + \frac{1}{2}[S_1,\Vod] \right)P_0 \ket{\nu,n}\label{eq:sw_mbot}\\
                    &= \delta_{mn}\delta_{\mu\nu}\left( m\omega_{\rm cl} + \mu\omega_{\rm qu} \right)\label{fwgewgee}\\
                    &\phantom{=}+ \frac{1}{2}\sum_{l=-\infty}^{\infty}\sum_{\lambda=0}^{\infty}
                    \Big(
                    \frac{P_0V^{ml}_{\mu\lambda}P_1 }{0+m\omega_{\rm cl}+\mu\omega_{\rm qu}-(U+l\omega_{\rm cl}+\lambda\omega_{\rm qu})}V^{ln}_{\lambda\nu}P_0
                    -P_0V^{ml}_{\mu\lambda} \frac{P_1V^{ln}_{\lambda\nu}P_0 }{U+l\omega_{\rm cl}+\lambda\omega_{\rm qu}-(n\omega_{\rm cl}+\nu\omega_{\rm qu})}
                    \Big)\nonumber\\
                    &= \delta_{mn}\delta_{\mu\nu}\left( m\omega_{\rm cl} + \mu\omega_{\rm qu} \right) - \frac{1}{2}\sum_{l=-\infty}^{\infty}\sum_{\lambda=0}^{\infty}
                    \left( \frac{P_0V^{ml}_{\mu\lambda}P_1V^{ln}_{\lambda\nu}P_0}{ U + (l-m) \omega_{\rm cl} + (\lambda-\mu) \omega_{\rm qu} } + \frac{P_0V^{ml}_{\mu\lambda}P_1V^{ln}_{\lambda\nu}P_0}{U + (l-n) \omega_{\rm cl} + (\lambda-\nu) \omega_{\rm qu}}\right).\label{fwwwlelee}
\end{align}
Inserting the expression~\eqref{Vqwer} for $V$, we further find
\begin{align}
  &P_0V^{ml}_{\mu\lambda}P_1V^{ln}_{\lambda\nu}P_0\nonumber\\
  = &P_0\sum_{\left\langle i,j \right\rangle\sigma}(-t_0)\im^{|m-l|+|\mu-\lambda|}J_{|m-l|}(g_l)j_{\mu,\lambda}(g_c)\chi_{ij}^{m-l}\xi_{ij}^{\mu-\lambda}c^{\dagger}_{i\sigma}c_{j\sigma}P_1\nonumber\\
  \cdot \,\,\,&P_1\sum_{\left\langle k,l \right\rangle\sigma'}(-t_0)\im^{|l-n|+|\lambda-\nu|}J_{|l-n|}(g_l)j_{\lambda,\nu}(g_c)\chi_{kl}^{l-n}\xi_{kl}^{\lambda-\nu}c^{\dagger}_{k\sigma'}c_{l\sigma'}P_0\\
  =&t_0^2\im^{|m-l|+|\mu-\lambda|+|l-n|+|\lambda-\nu|}J_{|m-l|}(g_l)j_{\mu,\lambda}(g_c)J_{|l-n|}(g_l)j_{\lambda,\nu}(g_c)
  \sum_{\left\langle i,j \right\rangle}\chi_{ij}^{m-l}\xi_{ij}^{\mu-\lambda}\chi_{ji}^{l-n}\xi_{ji}^{\lambda-\nu}\sum_{\sigma,\sigma'}P_0c^{\dagger}_{i\sigma}c_{j\sigma}c^{\dagger}_{j\sigma'}c_{i\sigma'}P_0.\label{eqb8}
  \end{align}
The last term is identified as the singlet projector, $\sum_{\sigma,\sigma'}P_0c^{\dagger}_{i\sigma}c_{j\sigma}c^{\dagger}_{j\sigma'}c_{i\sigma'}P_0=  2P^S_{ij}$.
Moreover, in the sum $\sum_{\left\langle i,j \right\rangle}$ over nearest neighbors each bond appears twice, once with $\chi_{ij}=\xi_{ij}=1$ and once with $\chi_{ij}=\xi_{ij}=-1$.
Using this fact, and the symmetry $\chi_{ij}=-\chi_{ji}$, $\xi_{ij}=-\xi_{ji}$, the sum can be rewritten as
\begin{align}
  \sum_{\left\langle i,j \right\rangle}\chi_{ij}^{m-l}\xi_{ij}^{\mu-\lambda}\chi_{ji}^{l-n}\xi_{ji}^{\lambda-\nu} 2P^S_{ij}
  =(-1)^{l-n+\lambda-\nu} \sum_{\left\langle i,j \right\rangle} \left( 1 + (-1)^{m-n+\mu-\nu} \right)P^S_{ij}.
\end{align}
In summary, the right hand side of Eq.~\eqref{eqb8} becomes
\begin{align}
  &t_0^2\im^{|m-l|+|\mu-\lambda|+|l-n|+|\lambda-\nu|}(-1)^{l-n+\lambda-\nu}J_{|m-l|}(g_l)j_{\mu,\lambda}(g_c)J_{|l-n|}(g_l)j_{\lambda,\nu}(g_c)\left( 1 + (-1)^{m-n+\mu-\nu} \right) \sum_{\left\langle i,j \right\rangle} P^S_{ij}.
\end{align}
Inserting this back into Eq.~\eqref{fwwwlelee}, and defining
\begin{align}
  \mathcal{J}^{mn}_{\mu\nu} = -&J_{\rm ex}\frac{1}{2}\sum_{l=-\infty}^{\infty}\sum_{\lambda=0}^{\infty}\im^{|m-l|+|\mu-\lambda|+|l-n|+|\lambda-\nu|}(-1)^{l-n+\lambda-\nu}J_{|m-l|}(g_l)j_{\mu,\lambda}(g_c)J_{|l-n|}(g_l)j_{\lambda,\nu}(g_c)\nonumber\\
  \cdot&\left( 1 + (-1)^{m-n+\mu-\nu} \right)\left( \frac{1}{1 + (l-m)\bar{\omega}_{\rm cl} + (\lambda-\mu)\bar{\omega}_{\rm qu}} + \frac{1}{1 + (l-n)\bar{\omega}_{\rm cl} + (\lambda-\nu)\bar{\omega}_{\rm qu}} \right),\label{eq:J_sp_driven}
\end{align}
we arrive at Eq.~\eqref{eq:mbot_spin_phot}.
\section{Connection between the Floquet-spin-photon Hamiltonian and resonant Raman-scattering}\label{sec:raman}
\begin{figure}
  \centering
  \includegraphics{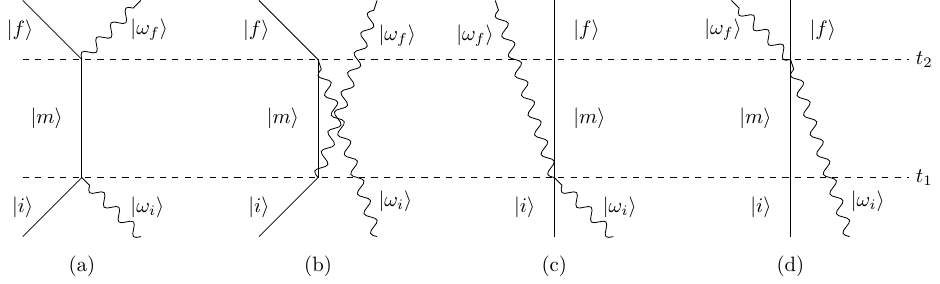}
  \caption{
    Sketches of the scattering processes from many particle state $\ket{i}\otimes\ket{\omega_i}$ to $\ket{f}\otimes\ket{\omega_f}$ for the time-dependent perturbation theory, which schematically representing Eqs.~\eqref{eq:tdepscat_begin}-\eqref{eq:tdepscat_end} for time-dependent perturbation theory~\cite{sakurai1967}.
    Since we are interested in second-order processes in $t_0$, the hopping operator $H_t$ acts twice at times $t_1$ and $t_2$, which changes the both the state of the solid and the state $\ket{\omega_{i/f}}$ of the field.
    }\label{fig:raman_processes}
\end{figure}

In this section we explain the connection between the Floquet-spin-photon Hamiltonian and the matrix elements of resonant Raman scattering.
Although the light matter Hamiltonian via the Peierls phase includes higher order nonlinear terms, a photon scattering with initial state $\ket{i}$ and final state $\ket{f}$ to the low-energy part of the Hilbert space can only occur via an intermediate state $\ket{m}$ with a charge excitation.
This is because the light-matter coupling only appears in the hopping term, which at half filling creates charge excitations in $\ket{m}$ when acting on any $\ket{i}$.
The Raman matrix elements can therefore be computed using time-dependent perturbation theory~\cite{sakurai1967} up to second order in $t_0$.
Within the second order time-dependent perturbation theory, the tunnelling operator $H_t$ acts twice and there are four distinct contributions $c_{(a)}$-$c_{(d)}$ to the amplitude  $c(t)=c_{(a)}+c_{(b)}+c_{(c)}+c_{(d)}$ for the transition from $\ket{i,\omega_i}$ to $\ket{f,\omega_f}$ within the time $t$;
the structure of these contribution is sketched in Fig.~\ref{fig:raman_processes}, and the algebraic expressions are given by
\begin{align}
  c_{(a)}(t) &= \int_0^t dt_2 \int_0^{t_2}dt_1 \sum_m \bra{f,\omega_f}H_t\ket{m}e^{i(\epsilon_f-\epsilon_m+\omega_f)t_2}\bra{m}H_t\ket{i,\omega_i}e^{i(\epsilon_m-(\epsilon_i+\omega_i))t_1}\label{eq:tdepscat_begin}\\
  c_{(b)}(t) &= \int_0^t dt_2 \int_0^{t_2}dt_1 \sum_m \bra{f,\omega_f}H_t\ket{m,\omega_f+\omega_i}e^{i(\epsilon_f-\epsilon_m-\omega_i)t_2}\bra{m,\omega_f+\omega_i}H_t\ket{i,\omega_i}e^{i(\epsilon_m+\omega_f-\epsilon_i)t_1}\\
  c_{(c)}(t) &= \int_0^t dt_2 \int_0^{t_2}dt_1 \sum_m \bra{f,\omega_f}H_t\ket{m,\omega_f}e^{i(\epsilon_f-\epsilon_m)t_2}\bra{m,\omega_f}H_t\ket{i,\omega_i}e^{i(\epsilon_m+\omega_f-\epsilon_i-\omega_i)t_1}\\
  c_{(d)}(t) &= \int_0^t dt_2 \int_0^{t_2}dt_1 \sum_m \bra{f,\omega_f}H_t\ket{m,\omega_i}e^{i(\epsilon_f+\omega_f-\epsilon_m-\omega_i)t_2}\bra{m,\omega_i}H_t\ket{i,\omega_i}e^{i(\epsilon_m-\epsilon_i)t_1}. \label{eq:tdepscat_end}
\end{align}
In all expressions, we can perform the $t_1$ integral in the form $\int_0^{t_2} dt_1 e^{ixt_1}\to -i e^{ixt_2}/x$, where we have neglect the contribution from the lower boundary of the $t_1$ integral, as it provides a contribution dependent on the turn-on of the interaction at $t=0$~\cite{sakurai1967}.
For the sum $c(t)=c_{(a)}+c_{(b)}+c_{(c)}+c_{(d)}$, this gives
\begin{align}
  c(t) &= -i \int_0^t dt_{2} \sum_m \tilde{M}(m) e^{i(\epsilon_f + \omega_f -(\epsilon_i+\omega_i))t_2}, 
\end{align}
where matrix elements and resolvents are summarized into
\begin{align}
  \tilde{M}(m) &= \frac{\bra{f,\omega_f}H_t\ket{m}        \bra{m}        H_t\ket{i,\omega_i}}{\epsilon_m-(\epsilon_i+\omega_i)}
                + \frac{\bra{f,\omega_f}H_t\ket{m,\omega_f+\omega_i}\bra{m,\omega_f+\omega_i}H_t\ket{i,\omega_i}}{\epsilon_m+\omega_f-\epsilon_i}\nonumber\\
               &+ \frac{\bra{f,\omega_f}H_t\ket{m,\omega_f}    \bra{m,\omega_f}    H_t\ket{i,\omega_i}}{\epsilon_m+\omega_f-\epsilon_i-\omega_i}
                + \frac{\bra{f,\omega_f}H_t\ket{m,\omega_i}    \bra{m,\omega_i}    H_t\ket{i,\omega_i}}{\epsilon_m-\epsilon_i}.
\end{align}
Evaluating the remaining integral and some algebra gives
\begin{align}
  |c(t)|^2/t &= \sum_{m,m'}\tilde{M}(m)\tilde{M}(m')^{*}
               \frac{\sin((\epsilon_f + \omega_f -(\epsilon_i+\omega_i))/2 \cdot t)^2}{t\left[(\epsilon_f + \omega_f -(\epsilon_i+\omega_i))/2\right]^2}
 \to  \pi \sum_{m,m'}\tilde{M}(m)\tilde{M}(m')^{*}\delta(\epsilon_f + \omega_f -(\epsilon_i+\omega_i))\label{eq:shell},
\end{align}
where in the second equation we have used the limit $\lim_{t\rightarrow\infty}\frac{\sin(xt)^2}{x^2t}=\pi \delta(x)$.
Imposing on shell condition for the energy, $\epsilon_f + \omega_f - \epsilon_i - \omega_i=0$, one can rewrite the matrix element
$\tilde{M}$ in the form
\begin{align}
  2\tilde{M} &= \bra{f,\omega_f}H_t\ket{m}\bra{m}\frac{H_t}{\epsilon_m-(\epsilon_i+\omega_i)}\ket{i,\omega_i}
               - \bra{f,\omega_f}\frac{H_t}{\epsilon_f+\omega_f - \epsilon_m}\ket{m}\bra{m}H_t\ket{i,\omega_i}\nonumber\\
             &+ \bra{f,\omega_f}H_t\ket{m,\omega_f+\omega_i}\bra{m,\omega_f+\omega_i}\frac{H_t}{\epsilon_m+\omega_f-\epsilon_i}\ket{i,\omega_i}
              - \bra{f,\omega_f}\frac{H_t}{\epsilon_f - \epsilon_m-\omega_i}\ket{m,\omega_f+\omega_i}\bra{m,\omega_f+\omega_i}H_t\ket{i,\omega_i}\nonumber\\
             &+ \bra{f,\omega_f}H_t\ket{m,\omega_f}\bra{m,\omega_f}\frac{H_t}{\epsilon_m+\omega_f-(\epsilon_i+\omega_i)}\ket{i,\omega_i}
              - \bra{f,\omega_f}\frac{H_t}{\epsilon_f + \omega_f-(\epsilon_m+\omega_f)}\ket{m,\omega_f}\bra{m,\omega_f}H_t\ket{i,\omega_i}\nonumber\\
             &+ \bra{f,\omega_f}H_t\ket{m,\omega_i}\bra{m,\omega_i}\frac{H_t}{\epsilon_m-\epsilon_i}\ket{i,\omega_i}
               - \bra{f,\omega_f}\frac{H_t}{\epsilon_f + \omega_f - (\epsilon_m + \omega_i)}\ket{m,\omega_i}\bra{m,\omega_i}H_t\ket{i,\omega_i}.
\end{align}
With the expression for the resolvent superoperator $\mathcal{L}(X)$ of the Schrieffer-Wolff transformation (Eq.~\eqref{eq:super_op}) and the partition of unity in the upper Hubbard band $I = \sum_{m}(\ket{m}\bra{m} + \ket{m,\omega_i}\bra{m,\omega_i} + \ket{m,\omega_f}\bra{m,\omega_f} + \ket{m,\omega_i+\omega_f}\bra{m,\omega_i+\omega_f})$, we can identify
\begin{align}               
\sum_m\tilde{M} &= \frac{1}{2}\bra{f,\omega_f} H_t \mathcal{L}(H_t)\ket{i,\omega_i} - \frac{1}{2}\bra{f,\omega_f}\mathcal{L}(H_t)H_t\ket{i,\omega_i}
             = \frac{1}{2}\bra{f,\omega_f} [H_t, \mathcal{L}(H_t)]\ket{i,\omega_i}.
\end{align}
With Eqs.~\eqref{eq:sw_mbot}~and~\eqref{eqs1generat}, the last term is identified as a matrix element of the (Floquet-) spin-photon Hamiltonian
\begin{align}  
\sum_m\tilde{M} = \bra{f,\omega_f} \tilde{H}_{\rm SP} \ket{i,\omega_i}.
\end{align}
The matrix element $\tilde{M}$ of resonant scattering on the energy shell in the low-energy sector is therefore exactly the same as that of the multi-block second-order effective low-energy Hamiltonian.\\
For the two modes present in the driven spin-photon Hamiltonian, we can therefore give a scattering interpretation to the matrix elements:
e.g., $\mathcal{J}^{01,10}\delta(\omega_{\rm qu}-\omega_{\rm cl})/\mathcal{J}^{10,01}\delta(\omega_{\rm qu}-\omega_{\rm cl})$ is the Raman scattering matrix element from the classical mode to the quantum mode or vice versa, $\mathcal{J}^{0n,10}\delta(\omega_{\rm qu-n\omega_{\rm cl}})$ are Hyper-Raman matrix elements and $\mathcal{J}^{11,\nu\nu}$ describes the dressed propagation of classical light through the sample.

\section{Explicit basis of the exact diagonalization}\label{sec:basis}
In order to reduce the Hilbert space, that is numerically diagonalized and explicitly ensure the exact preservation of the total parity throughout the numeric diagonalization, we transform the Hamiltonian (Eq.~\eqref{eq:HamED}) to parity diagonal basis, where it obtains a blockdiagonal structure.
Using the states
\begin{align}
  \sqrt{2}\ket{S} &= \left( c^{\dagger}_{1\uparrow}c^{\dagger}_{2\downarrow} - c^{\dagger}_{1\downarrow}c^{\dagger}_{2\uparrow} \right)\ket{0}\\
  \sqrt{2}\ket{T} &= \left( c^{\dagger}_{1\uparrow}c^{\dagger}_{2\downarrow} + c^{\dagger}_{1\downarrow}c^{\dagger}_{2\uparrow} \right)\ket{0}\\
  \sqrt{2}\ket{C^+} &= \left( c^{\dagger}_{1\uparrow}c^{\dagger}_{1\downarrow} + c^{\dagger}_{2\uparrow}c^{\dagger}_{2\downarrow} \right)\ket{0}\\
  \sqrt{2}\ket{C^-} &= \left( c^{\dagger}_{1\uparrow}c^{\dagger}_{1\downarrow} - c^{\dagger}_{2\uparrow}c^{\dagger}_{2\downarrow} \right)\ket{0},
\end{align}
the Hamiltonian of a single dimer becomes
\begin{align}
  H_{\mu\nu} &= \delta_{\mu\nu}\left(\mu\omega_\mathrm{c} \sum_{\psi=S,T} \ket{\psi}\bra{\psi} + \sum_{\psi=C^+,C^-} (U + \mu\omega_\mathrm{c})\ket{\psi}\bra{\psi}\right)\nonumber\\
         &-t_0\left( \im^{|\mu-\nu|}j_{\mu\nu}(g_{\rm cl})\xi_{12}^{\mu-\nu}\left( (1 + (-1)^{\mu-\nu})\ket{C^+}\bra{S} + (1 - (-1)^{\mu-\nu})\ket{C^-}\bra{S} \right) + \text{h.c.}\right)\label{eq:symmetrized_basis},
\end{align}
which is straightforwardly extended for multiple dimers.
The same basis can be used for the driven system in the Floquet frame, where the matrix elements of the kinetic term have to be extended for the Floquet sidebands.
One can see that the even charge excitation $\ket{C^+}$ and the odd charge excitation $\ket{C^-}$ couple to $\ket{S}$ only in combination with an even and odd change of the photon number, respectively.
This is a consequence of the conservation of total parity (electrons plus photons):
$\ket{S}$ and $\ket{C^+}$ are even, $\ket{C^-}$ is odd, and for the photon (which is included as an harmonic oscillator) the occupation number determines the parity.

\section{Derivation of the fourth-order effective Hamiltonian}\label{sec:FourthODer}

\subsection{Full expressions}\label{AppD1}
Using Eq.~\eqref{eq:eff_ham} as starting point, we can evaluate the fourth order effective Hamiltonian~\cite{Bravyi2011} first for the simplest choice of low-energy subspace, i.e., with no charge excitations and the quantum mode unoccupied (see Eq.~\eqref{eq:proj_empty}).\\
The zeroth order vanishes, while the second order gives
\begin{equation}
H_2 = J_{\mathrm{ex}}\sum_{\mu=0}^{\infty}\frac{j_{0,\mu}(g_{\rm qu})^2}{1 + \mu\bar{\omega}_{\rm qu}}\sum_{\langle i,j \rangle}\left( \vec{S}_i \vec{S}_j - \frac{1}{4} \right) = -J_{\rm ex}\sum_{\mu=0}^{\infty}\frac{j_{0,\mu}(g_{\rm qu})j_{\mu,0}(g_{\rm qu})}{1 + \mu\bar{\omega}}\sum_{\left\langle i,j \right\rangle}P^S_{ij},
\end{equation}
where we have used the usual form of the spin operators and the projector $P^s_{ij}$ on the singlet state on bond $ij$.\\
Let us next evaluate the three appearing fourth-order paths through the Hilbert space from Fig.~\ref{fig:pert_sketch}.
To abbreviate the notation, we split the contributions into a product of three factors:
the scalar contributions of the matrix elements $\mathcal{A}$, the energy resolvents $\mathcal{R}$ from the superoperator $\mathcal{L}$ and the operator contribution $\hat{O}$.
We start with the S1-path.
For the operator contribution, we obtain
\begin{equation}
  \hat{O}_{S1}= \sum_{\sigma_1,\sigma_2,\sigma_3,\sigma_4}P_0c^{\dagger}_{i_1\sigma_1}c_{j_1\sigma_1}c^{\dagger}_{j_1\sigma_2}c_{i_1\sigma_2}P_0c^{\dagger}_{i_2\sigma_3}c_{j_2\sigma_3}c^{\dagger}_{j_2\sigma_4}c_{i_2\sigma_4}P_0 = 4P_{i_1j_1}P_{i_2j_2},\label{OS1qqq}
\end{equation}
where we have used that the intermediate projectors $P_0$ constrain the position of the second and fourth hopping.
After expanding the nested commutator in Eq.~\eqref{eq:eff_ham} which gives the S1-path contribution, the matrix elements $\mathcal{A}$ and operator contributions $\hat{O}$ turn out to be the identical for all terms, and the resolvents can be summed up into
\begin{equation}
\mathcal{R}_{S1} = \frac{1}{U^3}\frac{8 + 4(\gamma + \alpha)\bar{\omega}_{\rm qu}}{(1 + \alpha\bar{\omega}_{\rm qu})^2(1 + \gamma\bar{\omega}_{\rm qu})^2}.
\end{equation}
Lastly, using the definition 
\begin{align}
\mathcal{J}^{\alpha\beta\gamma\delta\epsilon}(g_{\rm qu}) = j_{\alpha\beta}j_{\beta\gamma}j_{\gamma\delta}j_{\delta\epsilon}\im^{|\alpha-\beta|}\im^{|\beta-\gamma|}\im^{|\gamma-\delta|}\im^{|\delta-\epsilon|},
\end{align}
the matrix elements are given by
\begin{equation}
  \mathcal{A}_{S1} = t_0^4 j_{0\alpha}(g_{\rm qu})j_{\alpha 0}(g_{\rm qu}) j_{0\gamma}(g_{\rm qu})j_{\gamma 0}(g_{\rm qu}) = t_0^4(-1)^{\alpha+\gamma}\mathcal{J}^{0\alpha0\gamma0}.
\end{equation}
Combining the expressions for $\mathcal{A}_{S1}$, $O_{S1}$, and  $\mathcal{R}_{S1}$, we obtain
\begin{align}
  H_{S1} &= \frac{1}{8}\sum_{\substack{\left\langle i_1j_1 \right\rangle\\ \left\langle i_2j_2 \right\rangle}}\sum_{\alpha,\gamma=0}^{\infty}\mathcal{A}_{S1}\mathcal{R}_{S1}\hat{O}_{S1}\\
  &= K_0\sum_{\alpha,\gamma=0}^{\infty}\mathcal{J}^{0\alpha0\gamma0}\frac{(-1)^{\alpha+\gamma}(2 + (\gamma + \alpha)\bar{\omega}_{\rm qu})}{(1 + \alpha\bar{\omega}_{\rm qu})^2(1 + \gamma\bar{\omega}_{\rm qu})^2}\sum_{\substack{\left\langle i_1j_1 \right\rangle\\ \left\langle i_2j_2 \right\rangle}}P_{i_1j_1}P_{i_2j_2},
\end{align}
with $K_0=2t_0^4/U^3$, which gives Eq.~\eqref{eq:s1}.

The other two contributions describe processes with hopping inside the high-energy sector.
In $H_{\mathrm{NC}}$ we combine all processes with no intermediate charge excitation after two hoppings.
The fermionic hopping is therefore the same as for the processes contained in $H_{S1}$, but the sums and the resolvent are changed:
\begin{align}
  \hat{O}_{\mathrm{NC}}&= \sum_{\sigma_1,\sigma_2,\sigma_3,\sigma_4}P_0c^{\dagger}_{i_1\sigma_1}c_{j_1\sigma_1}c^{\dagger}_{j_1\sigma_2}c_{i_1\sigma_2}P_0c^{\dagger}_{i_2\sigma_3}c_{j_2\sigma_3}c^{\dagger}_{j_2\sigma_4}c_{i_2\sigma_4}P_0 = 4P_{i_1j_1}P_{i_2j_2}\label{ONCqqq}\\
\mathcal{R}_{\mathrm{NC}} &= -\frac{1}{U^3}\frac{1}{(1 + \alpha\bar{\omega}_{\rm qu})(\beta\bar{\omega}_{\rm qu})(1 + \gamma\bar{\omega}_{\rm qu})},\\
  \mathcal{A}_{\mathrm{NC}} &= t_0^4 \mathcal{J}^{0\alpha\beta\gamma0}\xi_{i_1j_1}^{0-\alpha}\xi_{j_1i_1}^{\alpha-\beta}\xi_{i_2j_2}^{\beta-\gamma}\xi_{j_2i_2}^{\gamma-0}
                  = t_0^4 \mathcal{J}^{0\alpha\beta\gamma0} (-1)^{\alpha+\gamma+\beta} \xi_{i_1j_1}^{\beta}\xi_{i_2j_2}^{\beta}
\end{align}
where in the last step we have used the symmetry $\xi_{ij}=-\xi_{ji}$.
When combining all terms,
\begin{align}
  H_{\mathrm{NC}} &= \sum_{\left\langle i_1j_1 \right\rangle}\sum_{\left\langle i_2j_2 \right\rangle}\sum_{\alpha,\beta,\gamma=0}^{\infty}(1-\delta_{\beta0})\mathcal{A}_{\mathrm{NC}}\mathcal{R}_{\mathrm{NC}}\hat{O}_{\mathrm{NC}},\label{efeher1123}
\end{align}
the expression contains the lattice sums $\sum_{\left\langle i_1,j_1 \right\rangle} P_{i_1j_1} \xi_{i_1j_1}^\beta$ (and similarly for $i_2,j_2$).
Because $\xi_{i_1j_1}$ is antisymmetric under exchange of $i_1$ and $j_1$,  $P_{i_1j_1}$ is symmetric, and each bond $\left\langle i_1,j_1 \right\rangle$ appears once in each direction, the sum contributes only for even $\beta$, where $\xi_{i_1j_1}^\beta=1$.
Hence the expression for $ H_{\mathrm{NC}}$ gives,
\begin{align}
  H_{\mathrm{NC}} 
  &  = -K_0\sum_{\alpha,\beta,\gamma=0}^{\infty}\mathcal{J}^{0\alpha\beta\gamma0}\frac{(1-\delta_{\beta0})(-1)^{\alpha+\beta+\gamma}(1 + (-1)^{\beta})}{(1 + \alpha\bar{\omega}_{\rm qu})(\beta\bar{\omega}_{\rm qu})(1 + \gamma\bar{\omega}_{\rm qu})}\sum_{\substack{\left\langle i_1j_1 \right\rangle\\ \left\langle i_2j_2 \right\rangle}}P_{i_1j_1}P_{i_2j_2},
\end{align}
where a factor $(1 + (-1)^{\beta})/2$ has been introduced to select the even $\beta$.

Finally, we derive the contribution from the DC path, which contain two intermediate charge excitations.
This requires the involvement of two separate bonds and is possible with two separate operator sequences:
The double occupation which is created first (on site $i_4$ in the expression below) can be the first ($\delta_1$) or the last one ($\delta_2$) to be broken up,
\begin{align}
  \hat{O}_{\mathrm{DC}}
  &=\sum_{\sigma_1,\sigma_2,\sigma_3,\sigma_4}P_0c^{\dagger}_{i_1\sigma_1}c_{j_1\sigma_1}c^{\dagger}_{i_2\sigma_2}c_{j_2\sigma_2}(\overbrace{\delta_{i_1j_3}\delta_{i_2j_4}\delta_{i_3j_1}\delta_{i_4j_2}}^{\delta_1} + \overbrace{\delta_{i_1j_4}\delta_{i_2j_3}\delta_{i_3j_2}\delta_{i_4j_1}}^{\delta_2})c^{\dagger}_{i_3\sigma_3}c_{j_3\sigma_3}c^{\dagger}_{i_4\sigma_4}c_{j_4\sigma_4}P_0\\
  &= 4(\delta_1 + \delta_2)P_{i_1j_1}P_{i_2j_2}\label{ODCqqq}
\end{align}
In the regular Hubbard model, these path contribute equally, but the cavity coupling introduces a phase which depends on the order.
The resolvents are given by
\begin{align}
\mathcal{R}_{\mathrm{DC}} &= -\frac{1}{U^3}\frac{1}{(1 + \alpha\bar{\omega}_{\rm qu})(2 + \beta\bar{\omega}_{\rm qu})(1 + \gamma\bar{\omega}_{\rm qu})}.
\end{align}
Combining the constraints $\delta_1$ and $\delta_2$ with the matrix elements gives 
\begin{align}
  (\delta_1+\delta_2)  \mathcal{A}_{\mathrm{DC}}
  &= t_0^4 \mathcal{J}^{0\alpha\beta\gamma0}\xi_{i_1j_1}^{0-\alpha}\xi_{i_2j_2}^{\alpha-\beta}(\delta_1+\delta_2)\xi_{i_3j_3}^{\beta-\gamma}\xi_{i_4j_4}^{\gamma-0}\\
  &=t_0^4 \mathcal{J}^{0\alpha\beta\gamma0}
                \Big(
                \delta_1  \xi_{i_1j_1}^{0-\alpha}\xi_{i_2j_2}^{\alpha-\beta}\xi_{j_1i_1}^{\beta-\gamma}\xi_{j_2i_2}^{\gamma-0}
                +
                \delta_2   \xi_{i_1j_1}^{0-\alpha}\xi_{i_2j_2}^{\alpha-\beta}\xi_{j_2i_2}^{\beta-\gamma}\xi_{j_1i_1}^{\gamma-0}
                \Big)\label{EQffd2}\\
  &=t_0^4 \mathcal{J}^{0\alpha\beta\gamma0}(-1)^{\beta}
               \Big(
                \delta_1  \xi_{i_1j_1}^{\beta-\alpha-\gamma}\xi_{i_2j_2}^{\alpha+\gamma-\beta}
                +
                \delta_2   \xi_{i_1j_1}^{\gamma-\alpha}\xi_{i_2j_2}^{\alpha-\gamma} 
                \Big),\label{EQffd3}
\end{align}
where in~\eqref{EQffd2} we have replaced the indices $i_3,j_3,i_4,j_4$ by $i_1,j_1,i_2,j_2$ according to the constraints $\delta_1$ and $\delta_2$, and in~\eqref{EQffd3} we have used the symmetry $\xi_{ij}=-\xi_{ji}$.
Following a similar argument as below Eq.~\eqref{efeher1123}, under the lattice sums $\sum_{\langle i_1j_1\rangle, \langle i_2,j_2\rangle}$ the term $\sim\delta_1$ and $\sim\delta_2$  contributes only if $(\alpha+\gamma-\beta)$ and $(\alpha-\gamma)$ are even, respectively.
Hence, under the lattice sum we can replace the matrix elements by
\begin{align}
(\delta_1+\delta_2)  \mathcal{A}_{\mathrm{DC}} \to t_0^4 \mathcal{J}^{0\alpha\beta\gamma0}(-1)^{\beta}\left(\delta_1 \frac{1 + (-1)^{\alpha+\beta+\gamma}}{2}+ \delta_2 \frac{1 + (-1)^{\alpha+\gamma}}{2} \right).\label{wfw;abdse;e}
\end{align}
Since summing over all bonds $\left\langle i_3j_3 \right\rangle, \left\langle i_4j_4 \right\rangle$ fulfills the conditions imposed by $\delta_1,\delta_2$ exactly one time each if $\left\langle i_1j_1 \right\rangle\neq \left\langle i_2j_2 \right\rangle$, the combination of all terms gives Eq.~\eqref{eq:DC} of the main text,
\begin{align}
  H_{\mathrm{DC}} &= \sum_{\substack{\left\langle i_1j_1 \right\rangle\\ \neq\left\langle i_2j_2 \right\rangle}}\sum_{\alpha,\beta,\gamma=0}^{\infty}\mathcal{A}_{\mathrm{DC}}\mathcal{R}_{\mathrm{DC}}\hat{O}_{\mathrm{DC}}
  = -K_0\sum_{\alpha,\beta,\gamma=0}^{\infty}\mathcal{J}^{0\alpha\beta\gamma0}(-1)^{\beta}\frac{2 + (-1)^{\alpha+\gamma}(1 + (-1)^{\beta})}{(1 + \alpha\bar{\omega}_{\rm qu})(2 + \beta\bar{\omega}_{\rm qu})(1 + \gamma\bar{\omega}_{\rm qu})}\sum_{\substack{\left\langle i_1j_1 \right\rangle\\ \neq\left\langle i_2j_2 \right\rangle}}P_{i_1j_1}P_{i_2j_2}.
\end{align}

\subsection{Leading order in $g_{\rm qu}$}\label{AppD2}
For further illustration, let us take these terms and expand in $g_{\rm qu}$ up to the fourth order.
\begin{align}
  H_{S1} &= \phantom{-}2K_0\left( 1 - g_{\rm qu}^2 \frac{\bar{\omega}_{\rm qu}(3 + 2\bar{\omega}_{\rm qu})}{(1 + \bar{\omega}_{\rm qu})^2} + g_{\rm qu}^4 \frac{\bar{\omega}_{\rm qu}^2(6 + 22\bar{\omega}_{\rm qu} + 25\bar{\omega}_{\rm qu}^2 + 8\bar{\omega}_{\rm qu}^3)}{(1+\bar{\omega}_{\rm qu})^3(1+2\bar{\omega}_{\rm qu})^2} \right)\sum_{\substack{\left\langle i_1j_1 \right\rangle\\ \left\langle i_2j_2 \right\rangle}}P_{i_1j_1}P_{i_2j_2}\\
  H_{DC} &= -2K_0\left( 1 - g_{\rm qu}^2 \frac{\bar{\omega}_{\rm qu}(3 + 2\bar{\omega}_{\rm qu})}{(1 + \bar{\omega}_{\rm qu})^2} + g_{\rm qu}^4 \frac{\bar{\omega}_{\rm qu}^2(7 + 25\bar{\omega}_{\rm qu} + 28\bar{\omega}_{\rm qu}^2 + 8\bar{\omega}_{\rm qu}^3)}{(1+\bar{\omega}_{\rm qu})^3(1+2\bar{\omega}_{\rm qu})^2} \right)\sum_{\substack{\left\langle i_1j_1 \right\rangle\\ \neq\left\langle i_2j_2 \right\rangle}}P_{i_1j_1}P_{i_2j_2}\\
  H_{NC} &= -2K_0g_{\rm qu}^4\frac{\bar{\omega}_{\rm qu} ^3}{(1+\bar{\omega}_{\rm qu})^2(1+2 \bar{\omega}_{\rm qu})^2}\sum_{\substack{\left\langle i_1j_1 \right\rangle\\ \left\langle i_2j_2 \right\rangle}}P_{i_1j_1}P_{i_2j_2}
\end{align}
By constraining $\left\langle i_2,j_2 \right\rangle$ to either the same bond or different bonds, we obtain the local fourth-order corrections to the Heisenberg exchange or the long-range mediated interactions:
\begin{align}
  J_{\rm loc } &=2K_0\left( 1 - g_{\rm qu}^2 \frac{\bar{\omega}_{\rm qu}(3 + 2\bar{\omega}_{\rm qu})}{(1 + \bar{\omega}_{\rm qu})^2}
                 + g_{\rm qu}^4  \frac{\bar{\omega}_{\rm qu}^2(6 + 21\bar{\omega}_{\rm qu} + 24\bar{\omega}_{\rm qu}^2 + 8\bar{\omega}_{\rm qu}^3)}{(1+\bar{\omega}_{\rm qu})^3(1+2\bar{\omega}_{\rm qu})^2} \right)\\
  K &= 2K_0g_{\rm qu}^4 \left( \underbrace{-\frac{\bar{\omega}_{\rm qu}^2(1 + 3\bar{\omega}_{\rm qu} + 3\bar{\omega}_{\rm qu}^2)}{(1+\bar{\omega}_{\rm qu})^3(1+2\bar{\omega}_{\rm qu})^2}}_{\kappa_{\rm S1}/2+\kappa_{\rm DC}/2} \underbrace{- \frac{\bar{\omega}_{\rm qu} ^2(\bar{\omega}_{\rm qu}+\bar{\omega}_{\rm qu}^2)}{(1+\bar{\omega}_{\rm qu})^3(1+2 \bar{\omega}_{\rm qu})^2}}_{\kappa_{\rm NC}/2} \right)\\
  &= -2K_0g_{\rm qu}^4 \frac{\bar{\omega}_{\rm qu}^2}{(1+\bar{\omega}_{\rm qu})^3}.
\end{align}
Note that to orders $g_{\rm qu}^0$ and $g_{\rm qu}^2$ $H_{S1}$ and $H_{DC}$ have contributions even when $\left\langle i_1,j_1 \right\rangle$ and $\left\langle i_2,j_2 \right\rangle$ are on different bonds.
This is because we employ the Schrieffer-Wolff transformation in a form which is not a linked cluster expansion.
When all terms are summed up, however, the contributions of order $g_{\rm qu}^0$ and $g_{\rm qu}^2$ to the interaction contributions cancel, so that leading contribution to the interaction is $\mathcal{O}(g_{\rm qu}^4)$.

\section{Generalization of the derivation of the interaction for cavity occupation and classical driving}\label{sec:fourthO_driven_deriv}
We can extend the perturbative scheme to account for occupation of the quantum mode by amending the energy resolvents and the amplitude terms.
Similarly in the Floquet frame the classical drive only changes the amplitudes and energies.
Since the effective Floquet Hamiltonian and the Floquet block Hamiltonian are translationally invariant under a global shift of the sideband index, it is sufficient to compute the effective Floquet Hamiltonian only for one sideband,
$n=0$.

Close to resonance the convergence radius of the perturbative series grows rapidly.
Higher order processes can therefore contribute more strongly, which formally limits our truncated scheme to small values of $t_0$~\cite{Bravyi2011}.
Since these resonances are also dependent on the light-matter coupling, the series convergence is not only controlled by $t_0$, but also the couplings $g_{\rm qu},g_{\rm cl}$, which helps our case.
Furthermore in physical systems, the frequencies $\omega_{\rm qu},\omega_{\rm cl}$ have a finite linewidth.
It is therefore reasonable to ignore resonances outside of the large cutoff we apply for the numerical evaluation of the interaction strength.\\
In second order we find
\begin{equation}
  H_{2,\nu} = J_{\mathrm{ex}}\sum_{m=-\infty}^{\infty}\sum_{\mu=0}^{\infty} \frac{j_{\nu,\mu}(g_{\rm qu})^2 J_{|m|}(g_{\rm cl})^2}{1 + (\mu-\nu)\omega_{\rm qu} + m\omega_{\rm cl}}\sum_{\left\langle i,j \right\rangle}\left( \vec{S}_i\vec{S}_j - \frac{1}{4} \right).
\end{equation}
To get the fourth-order terms, we use
\begin{align}
  \mathcal{J}^{\alpha\beta\gamma\delta\epsilon}_{abcde}(g_{\rm qu},g_{\rm cl}) = &j_{\alpha\beta}(g_{\rm qu})j_{\beta\gamma}(g_{\rm qu})j_{\gamma\delta}(g_{\rm qu})j_{\delta\epsilon}(g_{\rm qu})\cdot J_{|a-b|}(g_{\rm cl})J_{|b-c|}(g_{\rm cl})J_{|c-d|}(g_{\rm cl})J_{|d-e|}(g_{\rm cl})\nonumber\\
 \cdot&\im^{|\alpha-\beta| + |\beta-\gamma| + |\gamma-\delta| + |\delta-\epsilon|} \cdot \im^{|a-b| + |b-c| + |c-d| + |d-e|}.
\end{align}
The operator contribution $O_{S1}$, $O_{NC}$ and $O_{DC}$ to the three types of path is the same as for the empty cavity (Eqs.~\eqref{OS1qqq},~\eqref{ONCqqq}, and~\eqref{ODCqqq}).
The resolvent of the $S1$-term becomes
\begin{align}
  \mathcal{R}_{S1} &= \frac{8U + 4(\alpha+\gamma-2\nu)\omega_{\rm qu} + 4(a+c)\omega_{\rm cl}}{\left( U + (\alpha - \nu) \omega_{\rm qu} + a \omega_{\rm cl} \right)^2 \left( U + (\gamma - \nu) \omega_{\rm qu} + c \omega_{\rm cl} \right)^2}
\end{align}
and its amplitude
\begin{equation}
  \mathcal{A}_{S1} = t_0^4\mathcal{J}^{\nu\alpha\nu\gamma\nu}_{0a0c0}(g_{\rm qu},g_{\rm cl})(-1)^{\alpha+\gamma}(-1)^{a+c}.
\end{equation}
For the NC-term we find
\begin{equation}
  \mathcal{R}_{\mathrm{NC}}
  = \frac{1}{\left(U + (\alpha-\nu) \omega_{\rm qu} + a \omega_{\rm cl}\right) \left( (\beta-\nu) \omega_{\rm qu} + b \omega_{\rm cl} \right) \left( U + (\gamma-\nu) \omega_{\rm qu} + c \omega_{\rm cl} \right)}
\end{equation}
and
\begin{align}
  \mathcal{A}_{\mathrm{NC}} &= t_0^4 \mathcal{J}^{\nu\alpha\beta\gamma\nu}_{0abc0}(g_{\rm qu},g_{\rm cl})
                    (-1)^{\alpha+(\beta-\nu)+\gamma}(\xi_{i_1j_1}\xi_{i_2j_2})^{\beta-\nu}
                    (-1)^{a+b+c}(\chi_{i_1j_1}\chi_{i_2j_2})^{b}\\
                  &\to \frac{1}{2}t_0^4 \mathcal{J}^{\nu\alpha\beta\gamma\nu}_{0abc0}(g_{\rm qu},g_{\rm cl})(-1)^{\alpha+(\beta-\nu)+\gamma}(-1)^{a+b+c}
                    (1 + (-1)^{\beta-\nu+b}),
\end{align}
where the second equality hold only under the lattice sums, and we have used that the signs of the projected polarization can only either be aligned or anti-aligned for all bonds (see analogous to the arguments below Eq.~\eqref{efeher1123}).
Finally, the DC-term becomes
\begin{equation}
  \mathcal{R}_{\mathrm{DC}}
   = \frac{1}{\left( U + (\alpha-\nu) \omega_{\rm qu} + a \omega_{\rm cl} \right)\left( 2U + (\beta-\nu) \omega_{\rm qu} + b \omega_{\rm cl} \right)\left( U + (\gamma-\nu) \omega_{\rm qu} + c \omega_{\rm cl} \right)}
\end{equation}
and the matrix elements are (generalizing Eq.~\eqref{wfw;abdse;e}),
\begin{align}
  (\delta_1+\delta_2) \mathcal{A}_{\mathrm{DC}}
  &= t_0^4\mathcal{J}^{\nu\alpha\beta\gamma\nu}_{0abc0}(g_{\rm qu},g_{\rm cl})(\delta_1 + \delta_2)
    \xi_{i_1j_1}^{\nu-\alpha}\xi_{i_2j_2}^{\alpha-\beta}\xi_{i_3j_3}^{\beta-\gamma}\xi_{i_4j_4}^{\gamma-\nu}
    \chi_{i_1j_1}^{0-a}\chi_{i_2j_2}^{a-b}\chi_{i_3j_3}^{b-c}\chi_{i_4j_4}^{c-0}\\
  &\to \frac{1}{2}t_0^4\mathcal{J}^{\nu\alpha\beta\gamma\nu}_{0abc0}(g_{\rm qu},g_{\rm cl})(-1)^{\beta-\nu+b}
    ( 2 + (-1)^{\alpha+\gamma+a+c}(1 + (-1)^{(\beta-\nu)+b}) ),
\end{align}
where again the second equality holds under the lattice sums.
Combining all terms, we can write the effective Hamiltonian in the notation of the main text (Eqs.~\eqref{eq:s1} to~\eqref{eq:DC}), with modified matrix elements~\eqref{eq:s1path} to~\eqref{eq:DCpath}.
For the undriven cavity $g_{\rm cl}=0$ at nonzero cavity occupation $\nu$ we have
\begin{align}
  W^{\nu\alpha\beta\gamma\nu}_{S1}(g_{\rm cl},\bar{\omega}_{\rm qu}) &= \mathcal{J}^{\nu\alpha\beta\gamma\nu}(g_{\rm qu})\delta_{\beta,\nu}\frac{(-1)^{\alpha+\gamma}\left(2 + (\alpha+\gamma-2\nu)\bar{\omega}_{\rm qu}\right)}{\left( 1 + (\alpha - \nu) \bar{\omega}_{\rm qu} \right)^2 \left( 1 + (\gamma - \nu) \bar{\omega}_{\rm qu} \right)^2},\label{eq:W_S1}\\
  W^{\nu\alpha\beta\gamma\nu}_{NC}(g_{\rm cl},\bar{\omega}_{\rm qu}) &= \mathcal{J}^{\nu\alpha\beta\gamma\nu}(g_{\rm qu})(1-\delta_{\beta,\nu})\frac{(-1)^{\alpha+(\beta-\nu)+\gamma}
                                  (1 + (-1)^{\beta-\nu})}{\left(1 + (\alpha-\nu) \bar{\omega}_{\rm qu} \right) \left( (\beta-\nu) \bar{\omega}_{\rm qu} \right) \left( 1 + (\gamma-\nu) \bar{\omega}_{\rm qu} \right)},\label{eq:W_NC}\\
  W^{\nu\alpha\beta\gamma\nu}_{DC}(g_{\rm cl},\bar{\omega}_{\rm qu}) &= \mathcal{J}^{\nu\alpha\beta\gamma\nu}(g_{\rm qu})\frac{(-1)^{\beta-\nu}
                          ( 2 + (-1)^{\alpha+\gamma}(1 + (-1)^{(\beta-\nu)}) )}{\left( 1 + (\alpha-\nu) \bar{\omega}_{\rm qu} \right)\left( 2 + (\beta-\nu) \bar{\omega}_{\rm qu} \right)\left( 1 + (\gamma-\nu) \bar{\omega}_{\rm qu} \right)},\label{eq:W_DC}
\end{align}
while for the most general case of a driven cavity one has
\begin{align}
  W^{\nu\alpha\beta\gamma\nu;abc}_{S1} &= \mathcal{J}^{\nu\alpha\beta\gamma\nu}_{0abc0}(g_{\rm qu},g_{\rm cl})\delta_{\beta,0}\delta_{b,0}\frac{(-1)^{\alpha+\gamma}(-1)^{a+c}\left(2 + (\alpha+\gamma-2\nu)\bar{\omega}_{\rm qu} + (a+c)\bar{\omega}_{\rm cl}\right)}{\left( 1 + (\alpha - \nu) \bar{\omega}_{\rm qu} + a \bar{\omega}_{\rm cl} \right)^2 \left( 1 + (\gamma - \nu) \bar{\omega}_{\rm qu} + c \bar{\omega}_{\rm cl} \right)^2},\label{eq:W_S1_driven}\\
  W^{\nu\alpha\beta\gamma\nu;abc}_{NC} &= \mathcal{J}^{\nu\alpha\beta\gamma\nu}_{0abc0}(g_{\rm qu},g_{\rm cl})(1-\delta_{\beta,\nu}\delta_{b,0})\frac{(-1)^{\alpha+(\beta-\nu)+\gamma+a+b+c}
                          (1 + (-1)^{\beta-\nu+b})}{\left(1 \!+\! (\alpha\!-\!\nu) \bar{\omega}_{\rm qu} \!+\! a \bar{\omega}_{\rm cl}\right) \left( (\beta\!-\!\nu) \bar{\omega}_{\rm qu} \!+\! b \bar{\omega}_{\rm cl} \right) \left( 1 \!+\! (\gamma\!-\!\nu) \bar{\omega}_{\rm qu} \!+\! c \bar{\omega}_{\rm cl} \right)},\label{eq:W_NC_driven}\\
  W^{\nu\alpha\beta\gamma\nu;abc}_{DC} &= \mathcal{J}^{\nu\alpha\beta\gamma\nu}_{0abc0}(g_{\rm qu},g_{\rm cl})\frac{(-1)^{\beta-\nu+b}
                          ( 2 + (-1)^{\alpha+\gamma+a+c}(1 + (-1)^{(\beta-\nu)+b}) )}{\left( 1 + (\alpha-\nu) \bar{\omega}_{\rm qu} + a \bar{\omega}_{\rm cl} \right)\left( 2 + (\beta-\nu) \bar{\omega}_{\rm qu} + b \bar{\omega}_{\rm cl} \right)\left( 1 + (\gamma-\nu) \bar{\omega}_{\rm qu} + c \bar{\omega}_{\rm cl} \right)}.\label{eq:W_DC_driven}
\end{align}

\section{Interaction in the spin-photon-Floquet approach}\label{sec:drive_spin_phot_int}

To eliminate cavity and sideband fluctuations from an arbitrary cavity occupation, we can directly use
\begin{equation}
 H^{\nu}_{\mathrm{SP}} = H^{00,\nu\nu} - \sum_{b=-\infty}^{\infty}\sum_{\beta=0}^{\infty}\tilde{H}^{0b,\nu\beta}\frac{(1-\delta_{b0}\delta_{\nu,\beta})}{b\omega_{\rm cl} + (\beta-\nu)\omega_{\rm qu}}H^{b0,\beta\nu},
\end{equation}
since for fixed cavity occupation and sideband the unperturbed part of this elimination is proportional to the identity.
Using the same splitting up into an operator part $\hat{O}_{\mathrm{SP}}$, an amplitude $\mathcal{A}_{\mathrm{SP}}$ and a resolvent $\mathcal{R}_{\mathrm{SP}}$, we obtain
\begin{align}
  \hat{O}_{\mathrm{SP}} &= P_{i_1j_1}P_{i_2j_2}\label{eq:floq-spin-phot-opp}\\
  \mathcal{A}_{\mathrm{SP}} &= 2t_0^4\mathcal{J}^{\nu\alpha\beta\gamma\nu}_{0abc0}(g_{\rm qu},g_{\rm cl})(-1)^{a+b+c+(\alpha-\nu)+(\beta-\nu)+(\gamma-\nu)}(1+(-1)^{(\beta-\nu)+b})\\
  U^3((\beta-\nu)\bar{\omega}_{\rm qu} + b\bar{\omega}_{\rm cl})\cdot\mathcal{R}_{\mathrm{SP}} &= [(1+(\alpha-\nu)\bar{\omega}_{\rm qu} + a\bar{\omega}_{\rm cl})(1+(\gamma-\nu)\bar{\omega}_{\rm qu} + c\bar{\omega}_{\rm cl})]^{-1}\nonumber\\
                                  &+ [(1+(\alpha-\nu)\bar{\omega}_{\rm qu} + a\bar{\omega}_{\rm cl})(1+(\gamma-\beta)\bar{\omega}_{\rm qu} + (c-b)\bar{\omega}_{\rm cl})]^{-1}\nonumber\\
                                  &+ [(1+(\alpha-\beta)\bar{\omega}_{\rm qu} + (a-b)\bar{\omega}_{\rm cl})(1+(\gamma-\beta)\bar{\omega}_{\rm qu} + (c-b)\bar{\omega}_{\rm cl})]^{-1}\nonumber\\
                                  &+ [(1+(\alpha-\beta)\bar{\omega}_{\rm qu} + (a-b)\bar{\omega}_{\rm cl})(1+(\gamma-\nu)\bar{\omega}_{\rm qu} + c\bar{\omega}_{\rm cl})]^{-1}.\label{eq:floq-spin-phot-resolv}
\end{align}
Comparing this result to the NC-path (Eq.~\eqref{eq:W_NC_driven}), we find, that the first term of the spin-photon resolvent coincides with the NC-path resolvent.
The amplitudes and operator parts also coincide apart from a factor of 4, which comes from the four summands of the spin-photon resolvent.
Since we want to investigate the resonantly driven system (where the NC-terms dominate), let us expand $U^3\bar{\Delta}\cdot\mathcal{R}_{\mathrm{SP}}$ with the detuning $\bar{\Delta} = ((\beta-\nu)\bar{\omega}_{\rm qu} + b\bar{\omega}_{\rm cl})\ll 1$:
\begin{align}
  U^3\bar{\Delta}\cdot\mathcal{R}_{\mathrm{SP}} &= [(1+(\alpha-\nu)\bar{\omega}_{\rm qu} + a\bar{\omega}_{\rm cl}    )(1+(\gamma-\nu)\bar{\omega}_{\rm qu} + c\bar{\omega}_{\rm cl}    )]^{-1}\nonumber\\
                  &+ [(1+(\alpha-\nu)\bar{\omega}_{\rm qu} + a\bar{\omega}_{\rm cl}    )(1+(\gamma-\nu)\bar{\omega}_{\rm qu} + c\bar{\omega}_{\rm cl} - \bar{\Delta})]^{-1}\nonumber\\
                  &+ [(1+(\alpha-\nu)\bar{\omega}_{\rm qu} + a\bar{\omega}_{\rm cl} - \bar{\Delta})(1+(\gamma-\nu)\bar{\omega}_{\rm qu} + c\bar{\omega}_{\rm cl} - \bar{\Delta})]^{-1}\nonumber\\
                  &+ [(1+(\alpha-\nu)\bar{\omega}_{\rm qu} + a\bar{\omega}_{\rm cl} - \bar{\Delta})(1+(\gamma-\nu)\bar{\omega}_{\rm qu} + c\bar{\omega}_{\rm cl}    )]^{-1}\\
                  &= \frac{4}{(1+(\alpha-\nu)\bar{\omega}_{\rm qu} + a\bar{\omega}_{\rm cl})(1+(\gamma-\nu)\bar{\omega}_{\rm qu} + c\bar{\omega}_{\rm cl})}\nonumber\\
                  &+ 2\frac{21 + (\alpha+\gamma-2\nu)\bar{\omega}_{\rm qu} + (a+c)\bar{\omega}_{\rm cl}}{(1+(\alpha-\nu)\bar{\omega}_{\rm qu} + a\bar{\omega}_{\rm cl})^2(1+(\gamma-\nu)\bar{\omega}_{\rm qu} + c\bar{\omega}_{\rm cl})^2}\bar{\Delta} + \mathcal{O}(\bar{\Delta}^2)\label{eq:resolvent_detuning}
\end{align}
Under the assumption, that the drive does not introduce any additional charge-cavity-Floquet resonances, we can therefore control the validity of the spin-photon Hamiltonian approach by choosing a small detuning $\Delta \ll U$.
This is not too surprising, as in the computation on the Raman scattering we found, that for scattering on the energy shell, i.e.\ $\Delta=0$, the Floquet spin-photon Hamiltonian properly describes the scattering processes.
\end{document}